# An augmented swirling and round jet impinging on a heated flat plate and its flow and heat transfer characteristics

Premchand V Chandra[a,†], Pratikash Prakash Panda[b,a], Pradip Dutta[c,a]

[a] Interdisciplinary Centre for Energy Research (ICER), Indian Institute of Science, Bangalore-560012, India

[b] Department of Aerospace Engineering, Indian Institute of Science, Bangalore-560012, India

[c] Department of Mechanical Engineering, Indian Institute of Science, Bangalore-560012, India

[†] Author correspondence: premchandv@iisc.ac.in ; premsaf@gmail.com

**Abstract**

A geometrical mechanism that generates augmented swirling and round jets is being proposed. The proposed geometry has an axial inlet port and *3* tangential inlet ports, each of diameter *10mm*. A parameter called Split ratio, defined as the percentage of airflow split through these inlet ports, is introduced for the augmented jet. Flow at the split ratios (*SR-1, SR-2, SR-3,* and *SR-4*) results in a single augmented jet of swirling and round jets of diameter *D = 30mm*, for which impingement heat transfer is predicted using *3-D RANS* numerical simulations. Also, computations for conventional round jets and swirling jets generated by an in-house geometrical vane-swirler (at vane angles $\theta = 45^0, 60^0,$ and $30^0$) each of jet diameter *D = 30 mm* are performed for the Reynolds number (*Re = 6000 - 15,000*) and at a jet-plate distance (*H = 1.5D – 4D*). A comparative study of the flow structures for all the jets using computations is done, followed by a limited discussion on Particle Image velocimetry (*PIV*) flow visualization results. An impingement heat transfer analysis for all the jets is studied numerically. It is inferred that at a smaller jet-plate distance *H =1.5D* or *H/D =1.5*, the augmented jet and vane swirler jets showed an improved heat transfer from the impingement surface (heated flat plate) than at other *H/D*. In contrast, the conventional round jets showed maximum heat transfer at *H = 4D*. From the comparative study, the impingement heat transfer characteristics using the proposed augmented jet are better at an optimized jet-plate distance *H=1.5D* and at a split ratio (*SR-4*), with an enhancement in the average Nusselt number (*$Nu_{avg}$*) of *88%* than the conventional round jet and *101%* than the vane-swirler jet counterpart. Similarly, an enhancement in the stagnation Nusselt number (*$Nu_{stg}$*) of *189%* than the round jet is predicted for the proposed augmented jet at *SR-4*.

**Graphical abstract:**

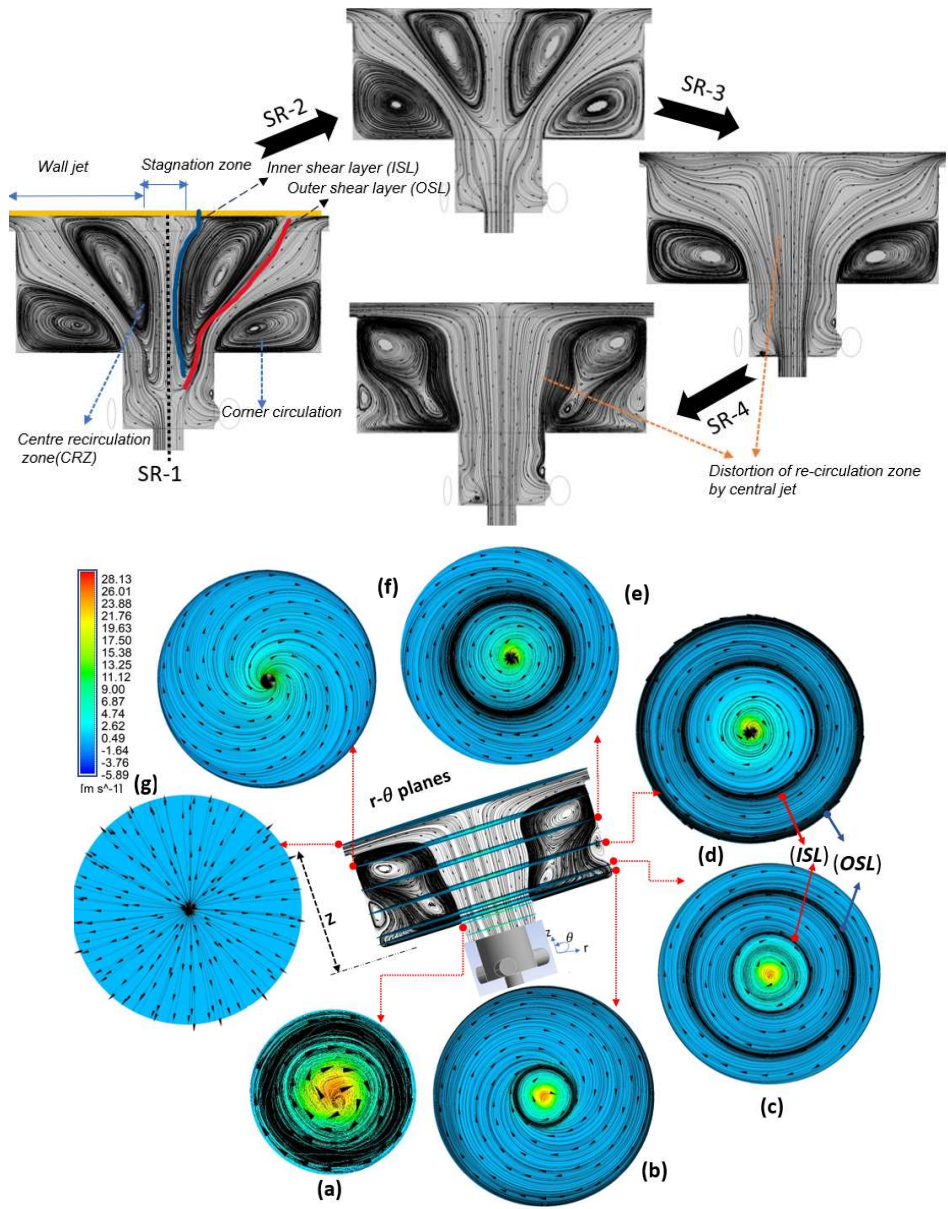

**Highlights**

- An augmented jet mechanism is proposed for enhancing impingement heat transfer.
- Predictive computational analysis using RANS models is carried out.
- Important parameters viz. jet-plate distance (H/D), Reynold number (Re), and geometrical Swirl no. (S) / Split ratio (SR – a new parameter introduced) is studied.
- To compare flow and heat transfer characteristics, a comparative study of swirling jets by geometrical vane swirler and conventional round jets is performed.
- A limited particle image Velocimetry (PIV) visualization study is carried out for the proposed augmented jet.

# 1. Introduction

Impingement heat transfer using conventional round (non-swirl) and swirling jets are efficient heat exchange methods that employ fluid jets impinging on the hot surface, which promises an enhanced heat transfer rate, because of which they find wide applications that include turbine blades and electronic cooling, etc. [29]. A single round jet impingement heat transfer characteristics depend mainly on the Reynolds number ($Re$) and dimensionless jet–plate distance ($H/D$) [3,16]. In contrast, a swirling jet is influenced by an additional parameter, the Swirl number ($S$), defined by the ratio of momentum fluxes in a typical swirl flow. Baughn et al. [2] studied a round turbulent air jet impinging on a heated flat plate at $H/D = 2 - 14$ for $Re$-23750, pointing out the existence of a secondary peak in Nusselt no. ($Nu$) at $H/D = 2$ and maximum stagnation point heat transfer ($Nu_{stg}$) at $H/D = 6$. Lee et al. [5] investigated the effect of nozzle diameter for a round jet impinging on a flat plate at jet-plate spacing ($H/D = 2 - 14$) for jet diameter ($D = 1.36 - 3.40$ cm) at $Re$-23000. They reported an increase in the stagnation Nusselt no. ($Nu_{stg}$) with a jet diameter ($D$). Cooper et al. [6] conducted flow field experiments and provided measurements on mean velocity and turbulence statistics for a single jet impinging for $Re$ - 23,000 & 70,000. The swirl flow with a radial uniformity of jet spread is a promising solution for enhanced impingement heat transfer [24]. From the swirl flow visualization studies, the characteristic tangential velocity components in a swirl flow cause the widening of the impingement and wall jet areas owing to its spiral-shaped motion, which can enhance the uniformity of jet spread. Ahmed et al. [17] carried out experimental and numerical investigation on swirling jet impingement and reported the highest convective heat transfer rate at low jet–plate distance ($H/D$). They also observed that stable recirculation zones were formed for a swirling jet at near–field impingement (i.e., low $H/D$), which positively affected the heat transfer coefficient. Nanan et al. [20] investigated the forced convective heat transfer aspects of swirl flows generated by twisted tapes of several twist ratios for different Reynolds no. ($Re = 4000$-$16000$) at $H/D = 2 – 8$ and reported maximum heat transfer at $H < 4D$. Beyond the impingement height $H > 4D$, there was an adverse heat transfer effect by swirling jets. The wide range of applications of impingement heat transfer attracted researchers to develop computational models that can predict the impingement flow physics and heat transfer characteristics within a reasonable computational cost. Due to the complex nature of the impinging flow, accurate prediction from computational methods is difficult as turbulence modelling becomes crucial. The time-dependent Direct numerical Simulation ($DNS$) and Large Eddy Simulation ($LES$) study revealed the vortical structures in a round impinging jet [15].

Dewan et al. [20] reviewed the numerical studies involving $DNS, LES,$ and hybrid $RANS/LES$. Though $LES$ and $DNS$ help in understanding the fundamental nature of flow and heat transfer by jet impingement, Reynolds Averaged Navier Stokes ($RANS$) models are suited for a first-place prediction of overall averaged heat transfer characteristics which help in the thermal system design. There is more than one turbulence $RANS$ model which predicts jet impingement heat transfer, and Zuckerman et al. [8] elucidated a qualitative review on the pros and cons of various $RANS$ turbulence models such as the $k-\varepsilon,\ k-\omega,$ Reynolds stress model, algebraic stress model, shear stress transport, and $v^2f$ model along with a list of empirical correlations for heat transfer.

### Nomenclature

| | | | |
|---|---|---|---|
| $Nu$ | Nusselt number | ***Superscripts*** | |
| $Re$ | Reynolds number | | |
| $Pr$ | Prandtl number | ' | Fluctuating terms in turbulence |
| $D$ | Jet diameter/ Diameter (mm) | + | Wall coordinates (non-dimensional) |
| $H/D$ | Dimensionless jet-plate distance | - | Time-averaged / mean (turbulence) |
| $r/D$ | Dimensionless radial distance | | |
| $k$ | Turbulence kinetic energy ($m^2s^{-2}$) | ***Subscripts*** | |
| $L$ | Length of potential core (mm) | | |
| $SR$ | Split ratio | avg | average |
| $S$ | Swirl number | eff | effective |
| $T$ | Temperature (K) | stg | stagnation |
| $u$ | Velocity (m/s) | max | maximum |
| $c_p$ | Specific heat at const. Pressure ($J\ kg^{-1}K^{-1}$) | min | minimum |
| $Pr_t$ | Turbulent Prandtl number | i, j | indices of coordinate direction |
| $C$ | Coefficients (constants) in RANS models | T | Turbulent |
| $P$ | Pressure (Pas or $N/m^2$) | L | Laminar |
| $P_k$ | Production Limiter | h | hub |
| $\theta$ | Vane / blade angle (in degree$^0$) | o | outer |
| x, y, z | Cartesian coordinates notation | ∞ | infinity |
| r, $\theta$, z | cylindrical coordinates notation | k | Turbulence kinetic energy |
| t | Blade thickness(mm) | $\omega$ | Specific rate of dissipation of kinetic energy |
| z | height (axial direction) | TOT | Total |
| ***Greek letters*** | | ***Abbreviations*** | |
| $\varepsilon$ | Rate of dissipation of TKE | RANS | Reynolds Averaged Navier Stokes |
| $\omega$ | Specific rate of dissipation of kinetic energy | CFD | Computational Fluid Dynamics |
| $\lambda$ | Thermal conductivity ($W\ m.K^{-1}$) | DNS | Direct Numerical Simulation |
| $\mu$ | Dynamic / turbulent viscosity ($kg\ m^{-1}\ s^{-1}$) | LES | Large Eddy Simulation |
| $\alpha$ | Eddy diffusivity ($m^2s^{-1}$) | TKE | Turbulence Kinetic Energy |
| $\beta$ | Closure coefficient in k-$\omega$ model | ISL | Inner Shear Layer |
| $\sigma$ | Turbulent Prandtl numbers | OSL | Outer Shear Layer |
| $\rho$ | Density ($kg.m^{-3}$) | ID | Inner Diameter |
| $\delta$ | Knocker delta from tensor algebra | TBL(BL) | Thermal( Boundary Layer) |
| $\nu$ | Kinematic/ turbulent viscosity ($m^2s^{-1}$) | SST | Shear Stress Transport |
| $\tau$ | Deviatoric stress tensor | SIMPLE | Semi Implicit Pressure Linked Equations |
| | | PIV | Particle Image Velocimetry |

R. Dutta et al. [11] conducted a comparative study on different $RANS$ turbulence models for predicting impingement heat transfer by turbulent slot jets. From the literature on jet impingement heat transfer, there is a mixed review on the heat transfer effectiveness of conventional round jets over swirling jet impingement and vice-versa. This work attempts to clarify the above ambiguity by a comparative numerical study using *3D RANS* simulations on round and swirling jets generated by vane swirlers at jet-plate distances (*H/D=1.5 – 4*) for Reynolds no. (*Re = 6500-15000*). We also propose a jet mechanism that augments a swirling and round jet into a single jet. Its impingement heat transfer characteristics are studied and compared with the conventional round and swirling jet generated by the vane swirlers.

The paper is structured to discuss the problem description in section 2, followed by the details of governing equations and turbulence models used in the present study in section 3. The computational domain and boundary conditions used are discussed in section 4. The predictive numerical simulation results and the flow and heat transfer characteristics of different jets are discussed in Section 6, and the conclusions are in Section 7.

## 2. Problem description

The flow structures for the conventional round and swirling jets impingement and for the proposed augmented jet are discussed, followed by a note on the application of impingement heat transfer in cavity receiver that forms the fundamental motivation for carrying out the present work.

### 2.1 Conventional Round and swirling jets:

The flow structure in a conventional round jet consists of a potential core region and a shear layer where the entrainment occurs. The Potential core is the distance between the exit of the nozzle to the point where the shear layer meets the centre of the jet, which usually ranges between *L = 4 - 6 D*, where *D* is the jet diameter. The impingement characteristics of a round jet have a stagnation regime which is a small region encompassing the stagnation point on the impingement surface, followed by an impingement regime formed by the jet impacting on the plate resulting in the change of mean flow direction vector from normal to radial direction which eventually extends to a wall jet regime, where the fluid sweeps the plate with a well-developed velocity profile (Fig.1). For a single jet impingement, the dimensionless nozzle-plate distance (*H/D*) which is essentially the distance between the jet orifice/nozzle outlet and impingement plate has an important effect on heat transfer characteristics [13,16]. When *H/D* is greater than the potential core, i.e., *H/D > L*, the Nusselt number is maximum at the stagnation point and is minimum for *H/D < L*. Swirl is a unique fluid flow feature used for many practical applications such as mixing, flame stabilization, cyclone separators, pneumatic conveyors, etc., due to its enhanced turbulence characteristics [24]. Introducing a swirl or rotational component to axisymmetric free jet results in a swirling jet.

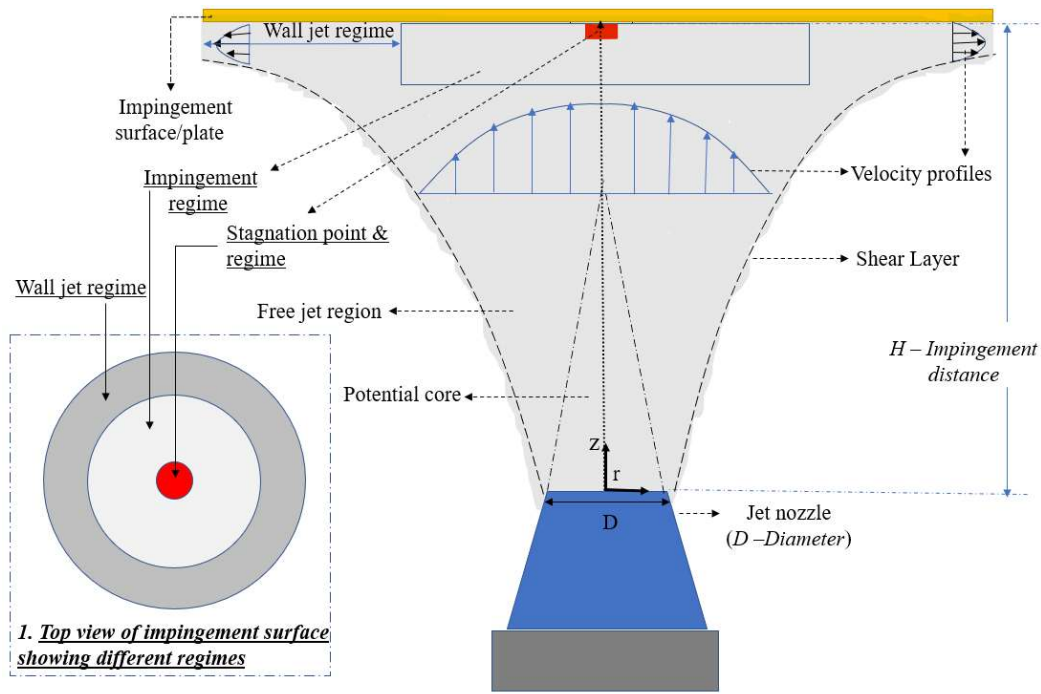

Fig 1. Typical flow structure and regimes in a conventional round jet impingement.

The swirl is strongest in the near field and decays in the far field. There are two shear layers in swirling jets: an Inner Shear Layer (*ISL*) and an Outer Shear Layer (*OSL*). The *ISL* demarcates the core region of the vortex flow from the mean flow field (refer Fig. 2. (b)). Vortex breakdown occurs in swirl flows when the tangential momentum exceeds the axial momentum. The characteristic of vortex breakdown is the transformation of the jet-like profile to a wake-like profile marked by a local minimum in velocity along the centreline axis of the jet due to an adverse pressure gradient. This causes a stagnation point (bubble-type) accompanied by a turbulent region of flow reversals downstream called the Recirculation zone, as shown in Fig.2. The intensity of the swirl is characterized by the Swirl number (*S*) defined as the ratio of the angular to axial momentum fluxes (axial component) which is independent of the method of generating swirl flow. Usually, the swirl no. *S= 0.48 - 0.94* is the range for the occurrence of vortex breakdown [4]. At the impingement surface, there is an enhancement in heat transfer due to the entrainment of fresh air at the Turbulent Boundary Layer (*TBL*), resulting in the breakup of large-scale eddies into intermittent and smaller eddies [24].

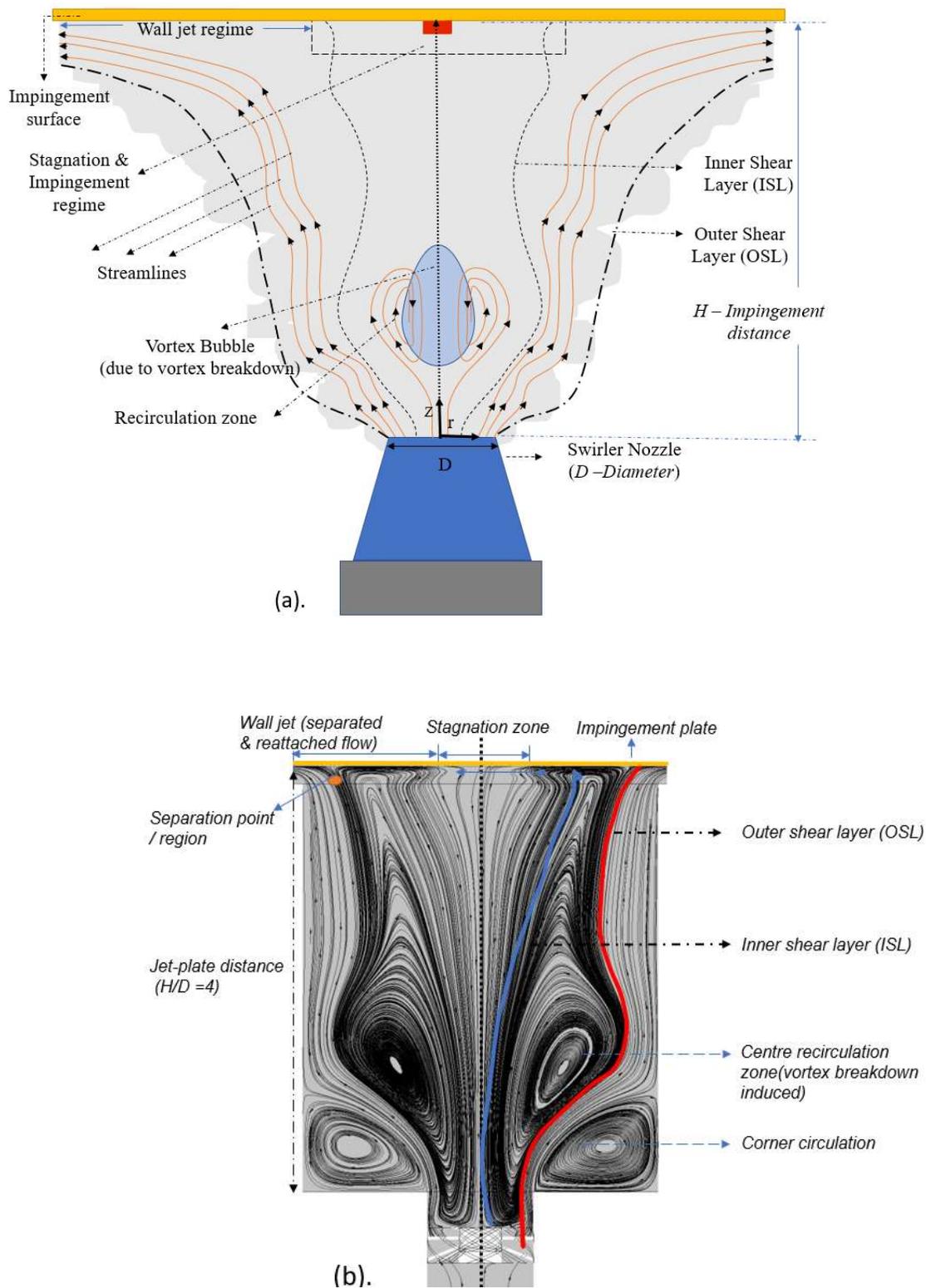

Fig 2. (a). Typical flow structure and regimes in a swirling jet impingement (b). Computed flow structures of impinging jet (at jet-plate distance *H/D = 4*) for a swirling jet generated by a *45˚* vane swirler at Reynolds no. (*Re- 12000*).

## 2.2 The proposed augmented jet (combined swirling and round jet)

A flow visualization study using Particle Image Velocimetry (*PIV*) is being carried out as the proposed augmented jet is a new flow structure of its kind, and hence scant in the literature is reported according to the author's knowledge. Fig.3. shows the averaged velocity and standard deviation contour plots from the *PIV* experiment. Further details from the computations and flow visualization experiments results are detailed in Section 6.1.

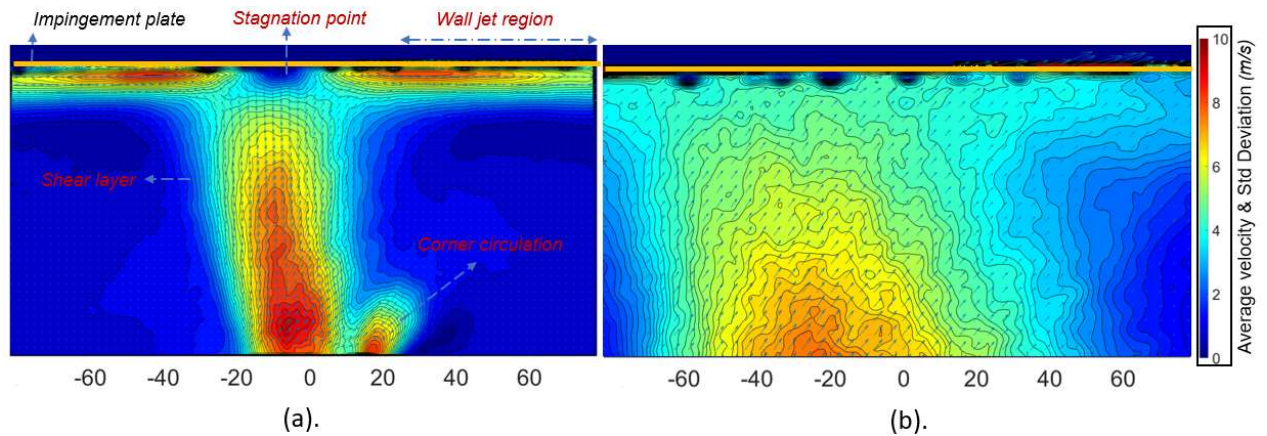

Fig.3. (a). Average velocity contour and streamline (b). Standard deviation contour plot from the Experimental *PIV* study for the proposed jet at mass flow (*500 SLPM*) at a split ratio *SR-4* (refer Table.1 in Section 4).

## 2.3 Cavity receiver Impingement heat transfer: An application perspective

A Cavity receiver is integral to Concentrated Solar Power (*CSP*) systems. A receiver is a collector geometry cum heat exchanger for solar thermal energy in a *CSP* thermal power plant. The air cavity receiver uses air as heat transfer fluid (*HTF*) which exchanges heat from the hot surfaces of the receiver by convection heat transfer mechanisms. Owing to the low thermal conductivity of air and limitations of heat exchange between the cavity receiver surfaces and the heat transfer fluid, the amount of heat recovered is limited. An enhanced heat transfer technique like impingement heat transfer is at stake for an improved rate of heat transfer from receiver surfaces which can enhance the thermal efficiency, thus resulting in maximum heat utilization from the solar radiation. The motivation for the present work is to apply such jet impingement mechanisms using the proposed augmented jet for effective heat transfer from the cavity receiver surface. The jet impingement mechanisms are targeted to exchange heat from the cavity receiver's back surface (focal plane surface) and lateral surfaces. Fig.4. shows a rough schematic of a *CSP* dish-receiver system.

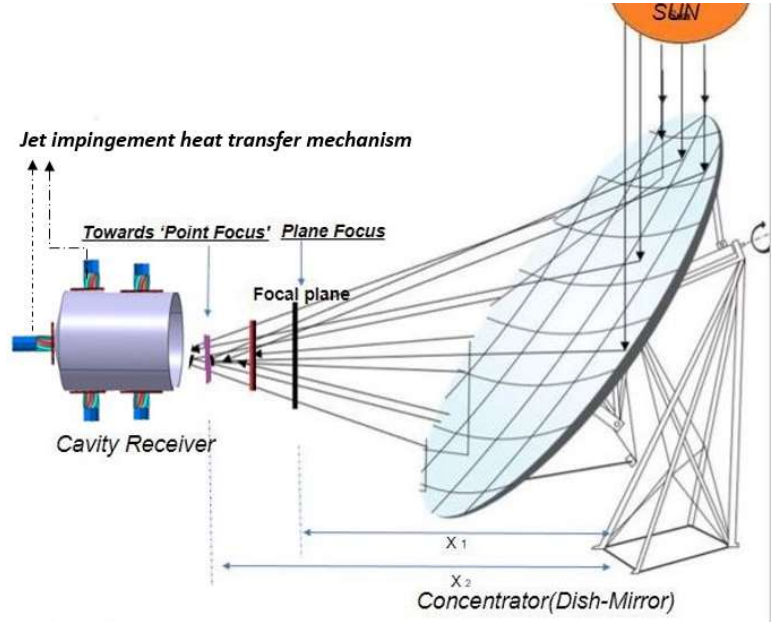

Fig.4. Schematic of *CSP* dish- receiver system with a Stand-alone concentrator dish (array of mirrors) and an air Cavity receiver with jet impingement mechanism for heat transfer.

## 3. Mathematical model

The Reynolds Averaged Navier-Stokes (*RANS*) equations for continuity, momentum, and energy are discussed, followed by turbulence models used for the present study in Appendix A

### 3.1. Governing equations

The Reynolds averaged continuity, momentum, and energy equations in the coordinate independent tensorial form are:

The averaged continuity equation,

$$\frac{\partial \bar{u}_i}{\partial x_i} = 0$$

(1)

The averaged momentum equation (*RANS*),

$$\rho \frac{\partial \bar{u}_i}{\partial t} + \rho \bar{u}_j \frac{\partial \bar{u}_i}{\partial x_j} = -\frac{\partial \bar{p}}{\partial x_i} + \frac{\partial}{\partial x_j}\left(2\mu \overline{S_{ij}} - \rho \overline{u'_i u'_j}\right)$$

(2)

The averaged energy equation,

$$\rho \frac{\partial \bar{T}}{\partial t} + \rho \bar{u}_j \frac{\partial \bar{T}}{\partial x_j} = \frac{\partial}{\partial x_j}\left(\frac{\lambda}{C_p}\frac{\partial \bar{T}}{\partial x_j} - \left(\rho \overline{\hat{u}_i T'}\right)\right)$$

(3)

The mean strain rate in equation (2) is given by:

$$\overline{S_{ij}} = \frac{1}{2}\left\{\frac{\partial u_i}{\partial x_j} + \frac{\partial u_j}{\partial x_i}\right\}$$  (4)

Where $C_p$, $\lambda$, and $\mu$ are the specific heat at constant pressure, thermal conductivity, and dynamic viscosity, respectively. The Reynolds stress term $\rho \overline{\hat{u}_i \hat{u}_j}$ in the averaged momentum equation (2) and turbulent heat flux term $\rho \overline{\hat{u}_i T'}$ in the averaged energy equation (3) must be defined by an appropriate turbulence model. The linear eddy viscosity model defines Reynold's stress which is given by,

$$-\rho \overline{\hat{u}_i \hat{u}_j} = 2\mu_t \overline{S_{ij}} - \frac{2}{3}\rho k \delta_{ij}$$  (5)

The turbulent heat flux is given by,

$$-\rho \overline{\hat{u}_i T'} = \frac{\mu_t}{Pr_t}\left(\frac{\partial \bar{T}}{\partial x_i}\right)$$

(6)

where, $k$ is the turbulent kinetic energy, $\mu_t$ is the turbulent or eddy viscosity and $Pr_t$ is the turbulent Prandtl number. The above equations are the RANS equations and modelling the Reynolds stress or turbulent stress term $\rho \overline{\hat{u}_i \hat{u}_j}$ requires a suitable eddy viscosity-based turbulence model to define $\mu_t$ in equations (5) and (6) to close the above governing equations. There are different linear viscosity-based turbulence models based on the equations solved for turbulent parameters such as turbulent kinetic energy ($k$), turbulent dissipation rate ($\varepsilon$), and specific dissipation rate ($\omega$). A detailed review of the appropriate RANS model that solves impingement heat transfer problems is elucidated by Zuckerman et.al. [8] and suggested that among all RANS models, SST hybrid 2 and 3 equation models such as SST $k - \omega$, SST Transition, and Transition $k$ - $k_l$ – $\omega$ had a relatively good predictive capability for jet impingement heat transfer. The details of the turbulence modelling can be referred in the Appendix. A

## 4. Computational domain and Boundary conditions

Fig.5. shows the computational geometries of the round jet, vane swirler, and the proposed augmented jet used in the present simulations. The round jet geometry has an inner diameter/jet diameter ($D=30$ mm) that opens to an expanding fluid domain and finally to an impingement plate separated by a variable distance $H/D = 1.5 - 4$, as shown in Fig. 5 (a). The geometrical vane swirler has an inner diameter (ID) of $D = 30mm$, which opens to swirler geometry containing 8 vanes/blades of angles ($\theta = 30°, 45°, 60°$ cases) that correspond to geometrical Swirl number (S) of *1.2, 0.7,* and *0.4* respectively calculated from,

$$S = \frac{2}{3}\left[\frac{1-\left(\frac{Dh}{D0}\right)^3}{1-\left(\frac{Dh}{D0}\right)^2}\right]\tan\theta , \qquad (7)$$

where, $\frac{Dh}{D0}$ is the ratio of hub diameter to the outer diameter of the swirler geometry and $\theta$ is the angle in degrees of flat vanes shown in Fig.5 (b). Due to blockage by the hub of the swirler, the Reynolds number (Re) calculated for this case is given by,

$$Re = \frac{\rho v (D_s - D_h)}{\mu}$$

(8)

Where, $D_s$ is the diameter of the swirler (equal to jet diameter $D = 30mm$), $D_h$ is the diameter of the hub, $\rho$ is the density, $\mu$ is the dynamic viscosity and $v$ is the inlet bulk velocity of the fluid.

Fig. 5 (c). is the proposed augmented geometry which has a total of *4* inlet ports, each of diameter *10mm*, of which *3* ports are tangential (aligned *120°* apart) and an axial inlet port. The total mass flow rate (is split through these *4* inlets. Table 1 shows the details of different split ratios for which simulations are performed. Both structured and unstructured meshes using *ICEM* meshing software and *ANSYS* meshing workbench are used for the present computations. The numerical simulations are carried out using a pressure-based solver with velocity inlet boundary conditions, calculated from the inlet Reynold number (Re) assuming a fully developed flow at the inlet. A constant heat flux on the impingement plate is applied as the thermal boundary condition. As the computational geometry is variable with the jet-plate distance (H/D), the number of mesh elements and nodes is also a variable.

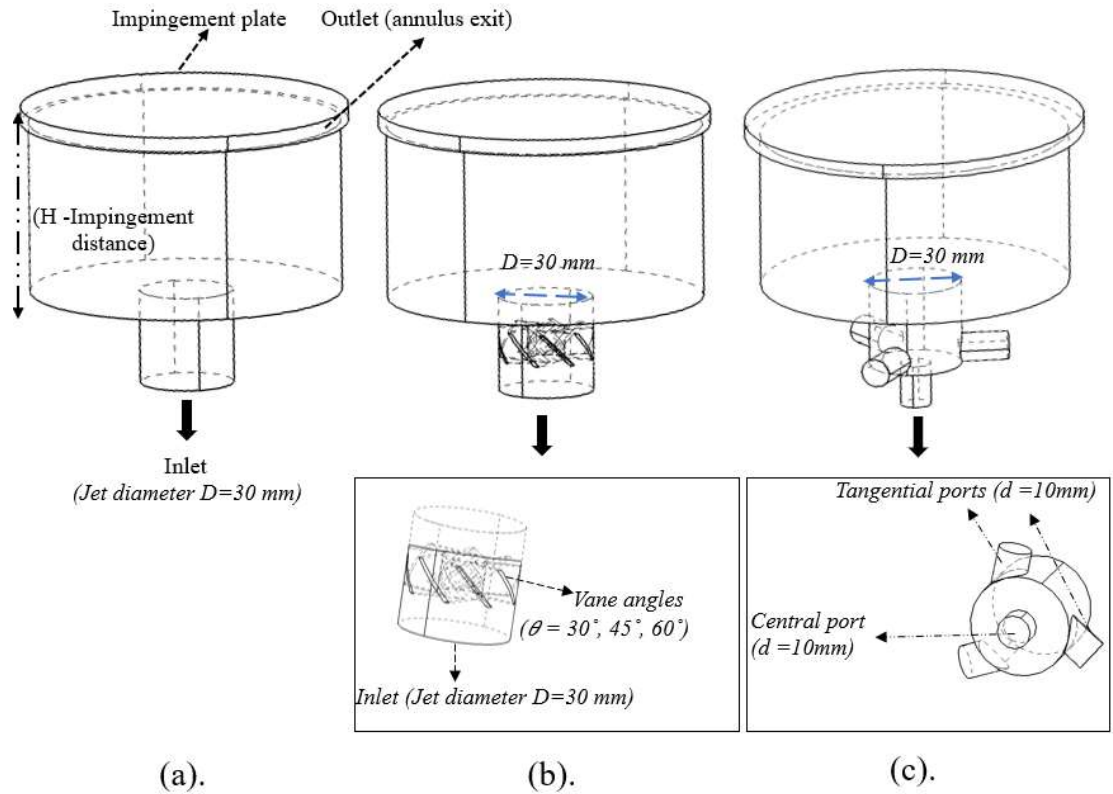

Fig.5. Geometries used in the present study (a). Round jet (b). Vane swirler jet (c). Aerodynamic swirler jet.

Table 1:    Details of split ratios and the percentage of flow through different ports.

| S.No | Split Ratios | Percentage of flow split through axial port (central jet) $\dot{m}_C$ | Percentage of flow split through 3 tangential ports $\dot{m}_T$ |
|---|---|---|---|
| 1 | SR-1 | 10% | 90% |
| 2 | SR-2 | 25% | 75% |
| 3 | SR-3 | 40% | 60% |
| 4 | SR-4 | 50% | 50% |

A grid-independent study using $4$ different mesh sizes based on the length of the finite volume ($FV$) element is carried out. Fig.6.(a) shows the grid-independent plot where the $Mesh-1$, $Mesh-2$, $Mesh-3$, and $Mesh-4$ corresponds to a $FV$ element of base length $1.2 \times 10^{-3} m$, $1 \times 10^{-3} m$, $8 \times 10^{-4} m$, and $5 \times 10^{-4} m$ respectively except at the grid refinement zones such as the boundary layer ($BL$) region near the impingement plate where a fine mesh is a mandatory requirement, shown in Fig.6 (b). Hence a first layer thickness at this grid refinement region is chosen to be $1 \times 10^{-4} m$ in the present computations, which satisfies the wall $y+$ criteria. The *Mesh-2* is chosen as grid independent candidate from the grid independent study.

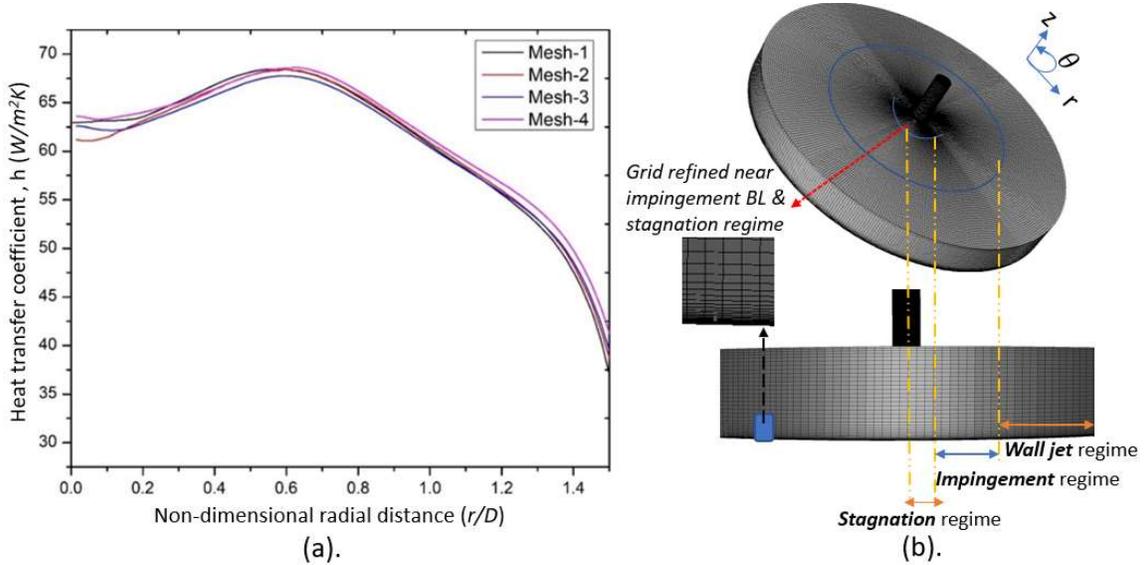

Fig.6 (a). Grid-independence study using different mesh sizes (b). The meshed *CFD* domain indicating the grid refinement near stagnation regimes and impingement Boundary layer (*BL*).

## 5. Solution methodology and validation of turbulence model

The problem is solved for continuity, momentum, and energy equations in *r, θ ,* and *z* directions along with transport equations for Transition $k$ - $k_l$ - $\omega$ and *SST* $k - \omega$ turbulence models of *RANS* detailed in Appendix A.1 & A.2. All the numerical computations are performed using cell centred finite volume code *ANSYS* Fluent *21.1*. The solution methodology involves coupling pressure-velocity using the Semi-implicit method for Pressure Linked Equations (*SIMPLE*) scheme with least square cell-based gradient spatial discretization. Second order upwind scheme for discretization is applied for all momentum, turbulent kinetic energy, laminar kinetic energy, pressure, specific dissipation rate equations, and *PRESTO* (Pressure Staggering option) scheme is used for pressure discretization applicable for swirl dominated flows. The normalized residuals are set to $10^{-7}$ for all the variables as the minimum condition for convergence. To validate the computational model used in the present study, the computational results of swirl flow velocity profiles and impingement heat transfer in terms of Nusselt no. (Nu) are compared with the PIV experimental data of R. Gopakumar et al. [4] and Baughn et al. [2], respectively. Fig.7. shows the validation plots for axial and azimuthal velocities of a free swirl flow for a *45˚* vane swirler case at *Re – 6500* and at a height *z = 7mm* from the swirler dump plane.

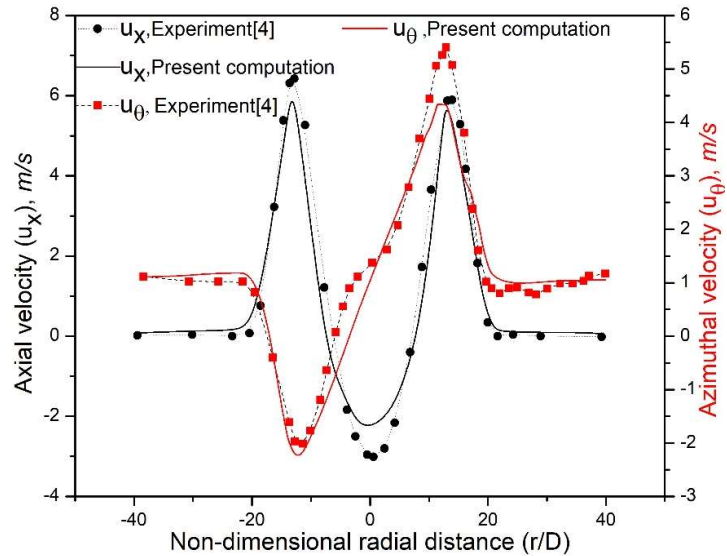

fig.7. Computed axial and azimuthal velocity profiles for a *45˚* swirler at *z = 7 mm* from the swirler exit/dump plane compared with *PIV* experiment of [4].

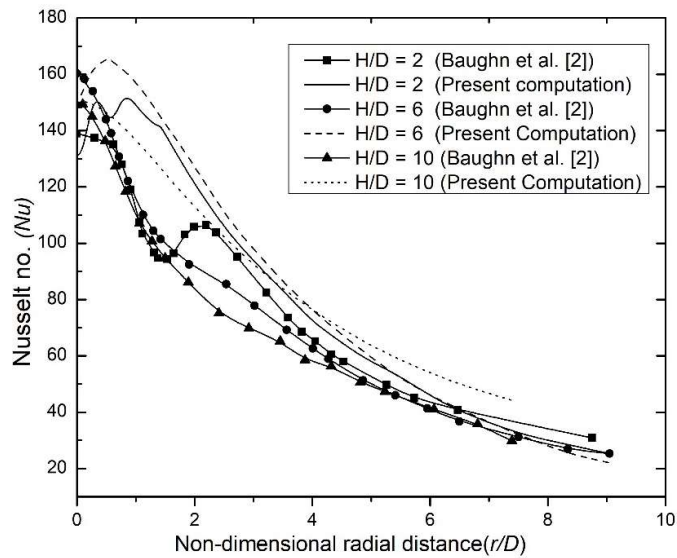

Fig. 8. Experimental [2] and computed Nusselt number distribution at different jet-plate distance at Reynolds number (*Re- 23,750*).

The validation plots for Nusselt number distribution for round jets at different jet-plate distances (*H/D = 2, 6, & 10*) and Reynolds no. (*Re – 23,750*) is shown in Fig.8. The deviation in the computed velocity profiles from the experimental data is acceptable as the turbulent experiment data is compared with the results from *RANS* Turbulence modelling, which is less accurate than the celebrated *LES* and *DNS* simulations. It can be seen from the plot that at the stagnation point (*r/D = 0*) due to grid refinement using a fine mesh size of $1 \times 10^{-4} m$ for the first layer thickness (*BL* near the impingement plate, Fig.6 (b)), there is a good agreement of computed $Nu_{stg}$ (stagnation Nusselt no.) with experiments of Baughn et al. [2]. The computed heat transfer near the wall jet (*r/D >2.5*) is also reasonably agreeing with the experiments. Whereas near the impingement regime (*0 > r/D > 2.5*) the deviation is prevalent, which may be attributed due two reasons: (a). Limitations with the fine meshing and grid refinement in the impingement regime, as fine mesh sizes, imposes additional computational cost for the *RANS* simulation. (b). Again, the limitations posed by *RANS* turbulence models for jet impingement physics, as *RANS* uses averaged equations best suits for studying an average flow and heat transfer phenomena. However, the errors are within acceptable limits as detailed by Zuckerman et al. [8].

## 6. Results and Discussions

This section discusses the jet impingement characteristics for the proposed augmented jet from 3D - RANS simulations along with limited experimental PIV results, followed by swirling jet (vane swirlers) and conventional round jet impingement. Parameters that influence convective heat transfer viz. Reynolds number (*Re*), jet-plate distance (*H/D*), geometrical swirl number (*S*) corresponding to vane angle (*θ*) and Split ratio (*SR*), a new parameter introduced for the study of the proposed augmented jet, are carried out. Finally, we discuss the comparative analysis of impingement heat transfer for all the jets and their optimized conditions for maximum heat transfer.

### 6.1. Proposed augmented jet

The effect of Split ratio (*SR*), a new parameter for the augmented jet on the impingement heat transfer, is discussed along with other parameters such as Reynolds no. (*Re*) and jet-plate distance (*H/D*). Also presented are the flow field in terms of average velocity contour plot with streamlines for the orthogonal front (*r-z* plane) and top (*r- θ* plane) planes alongside a limited PIV experimental result and, finally, the temperature contour plot on the impingement plate at different split ratios (*SR*).

### 6.1.1. Effect of Split ratio (SR)

The split ratio, as dealt with in section 4, is the percentage of flow split through the central port (for an axial jet) and equally through 3 other tangential ports, each of diameter $D_p$ = *10mm* adding up to mass flow rate equivalent to that of a round jet of *D = 30mm* for comparing the heat transfer characteristics with round jets. Fig.9. is the flow structures (streamlines) and their evolution at different split ratios. At split ratios *SR-1* and *SR-2*, the flow resembles typical swirling flow structures shown in Fig. 2. (b). At higher split ratios *SR-3* and *SR-4*, the percentage of flow through central jet counterparts increases, thus resulting in distortion of re-circulation zones, and the resulting flow feature is no more a predominant swirling jet but an augmented swirl and round jet. The results from the PIV flow visualization experiments also verify these, as shown in Fig.10. (a). With the increase in split ratio at *SR-3* and *SR-4*, the stagnation zone tends to a stagnation point which may positively impact the heat transfer at the stagnation regime. The swirl is strong at the SR-1 near the stagnation and impingement regime, as shown in Fig.10. (b) and it decreases at SR-4. Fig.11. (a) shows the local Nusselt number distribution at different split ratios (refer Table.1) for a Reynolds number (*Re-12000*) at a jet-plate distance $H/D = 2$. The maximum heat transfer characteristic is at split ratio *SR-4* with an average Nusselt number $Nu_{avg} = 188.82$ followed by the split ratios *SR-3* ($Nu_{avg} = 82.80$), *SR-1* ($Nu_{avg} = 57$) and *SR-2* ($Nu_{avg} = 64.70$). Hence, *SR-4* is the optimal split ratio corresponding to maximum impingement heat transfer characteristics for the proposed augmented jet. Fig.11. (b) shows the influence of Reynolds no. ($Re = 6500 - 15000$) on the local Nusselt no. distribution at a jet-plate distance ($H/D = 2$) corresponding to the split ratio (*SR-4*), which indicates an increase in heat transfer with Reynolds number, a trivial phenomenon observed for all jets [2-3, 20].

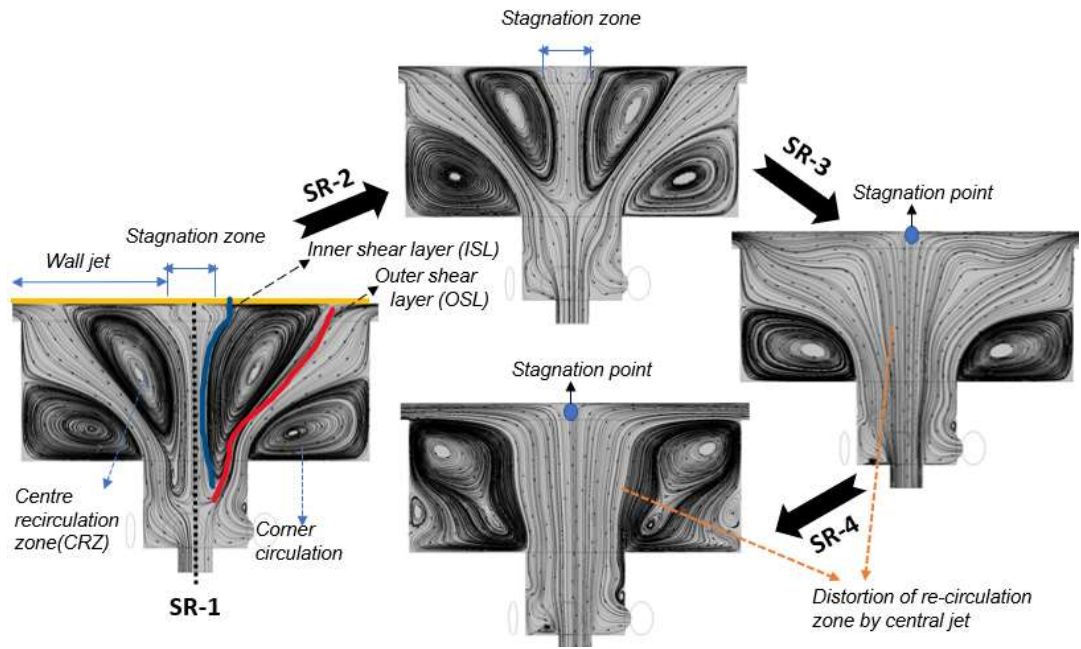

Fig.9. Computed streamlines at different split ratios from *SR-1* to *SR-4* showing the distortion of swirl flow by the central round jet.

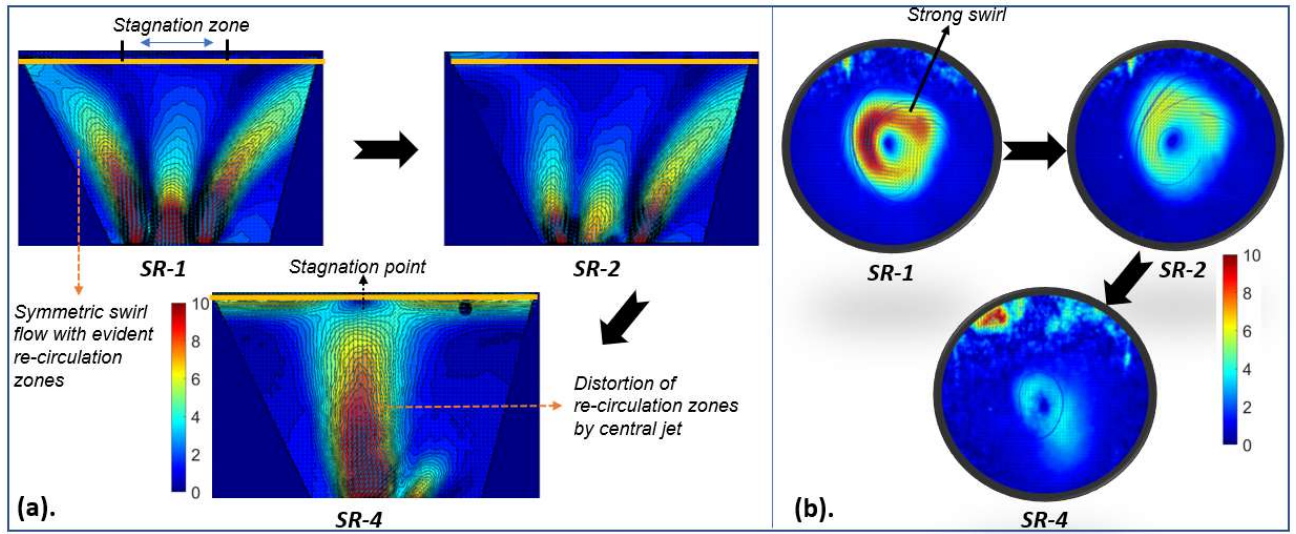

Fig.10. Averaged velocity contour and streamline plots from the PIV experiments for the proposed augmented jet impinging on a plate ($H/D = 4$) at split ratios *SR-1, SR-2,* and *SR-4* corresponding to a total mass flow ($\dot{m} = 500$ *SLPM*). (a). At orthogonal front ($r$-$z$) plane (b). At an $r$-$\theta$ plane ($H/D = 4$).

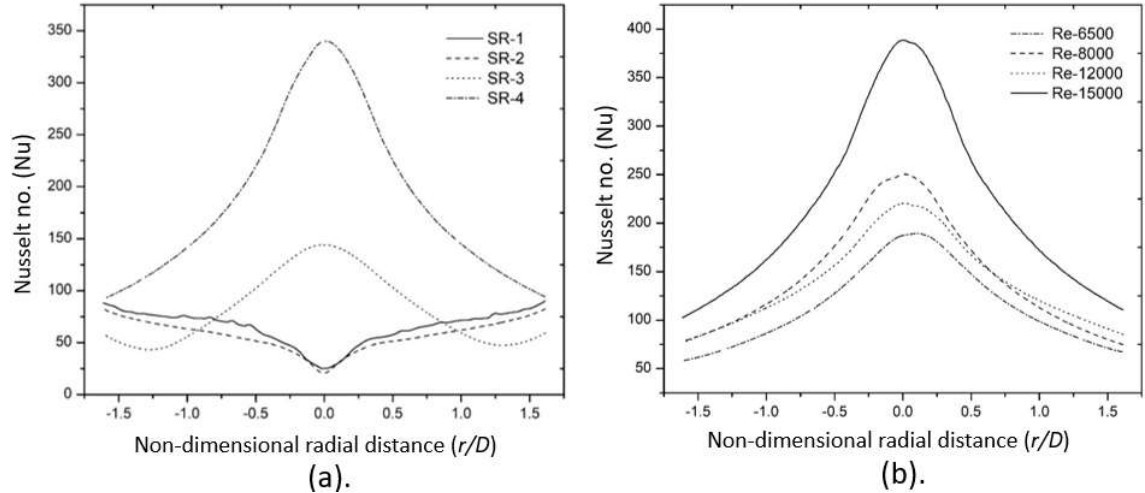

Fig.11. (a). Effect of split ratio for the proposed augmented jet at *Re-12000*, and *H/D =2*.
(b). Effect of Reynolds no. (Re) for the proposed augmented jet at *SR-4*.

### 6.1.2. Effect of jet-plate distance (H/D)

Fig.12. shows the Nusselt number distribution at different jet-plate impingement distances ($H/D = 1.5, 2, 3,$ and $4$) at *SR-4* split ratio (which is optimum) for *Re-12000*. The maximum heat transfer characteristic is at the lowest jet-plate distance $H/D = 1.5$, with an average Nusselt

no. ($Nu_{avg}$ = *140.5*) and the stagnation Nusselt no. ($Nu_{stg}$ = *220*). The heat transfer decreases with increased *H/D*, like swirl jet impingement, as discussed in section 6.2. The average Nusselt no. are $Nu_{stg}$ =*127.16*, *75.44*, and *59.3* for *H/D* = *2,3,* and *4,* respectively.

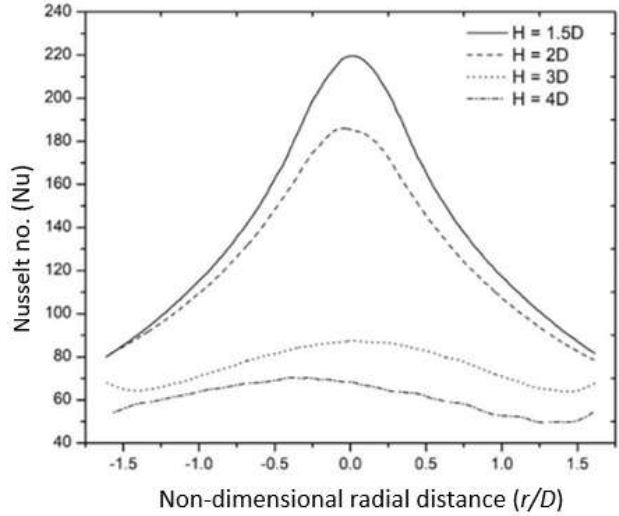

Fig.12. Effect of jet-plate distance (*H/D*) for the proposed jet at *SR-4* for *Re-12000*.

### 6.1.3   Flow field and temperature contours

From Fig.13 (a), (b), which corresponds to split ratios *S-1* & *S-2,* there is a stable swirl jet impingement feature akin to Fig.2. (b). At the higher split ratios *SR-3* & *SR-4,* the axial momentum from the central jet flow is high enough to distort the swirl flow resulting in a reduction of the stagnation zone, which can enhance heat transfer near the stagnation zones. Fig. 14. shows the average velocity contours and streamlines at *SR-4* for *Re-12000* at various *r- θ* planes. At (a) *z = 10mm, (b) 15mm,* and *(f) 50mm*, the streamlines indicate a swirl component, and the streamlines are radially outward at *z = 60mm*, meaning the impingement plane. The growths of the inner shear layer (*ISL*) and outer shear layers (*OSL*) are shown at (c) *z = 20mm* and (d) *z = 30mm*, beyond which there is only one shear layer, as shown in (e). *z = 40mm*, which may indicate distortion and merge of *ISL* and *OSL* due to high axial momentum by the central jet at this Split ratio (*SR-4*) (refer to Fig. 9). However, high-fidelity time-dependent *LES or DNS* simulations and flow visualization experiments are needed to support the above reasonings. Fig.13. (d). shows the temperature contours (at SR-*4*) with zones Z-*1,* Z-*2,* and Z-*3,* which corresponds to temperature distribution at stagnation, impingement, and wall jet regimes, indicating better heat transfer characteristics at *Z-1 and Z-2* than at wall jet regime *Z-3*.

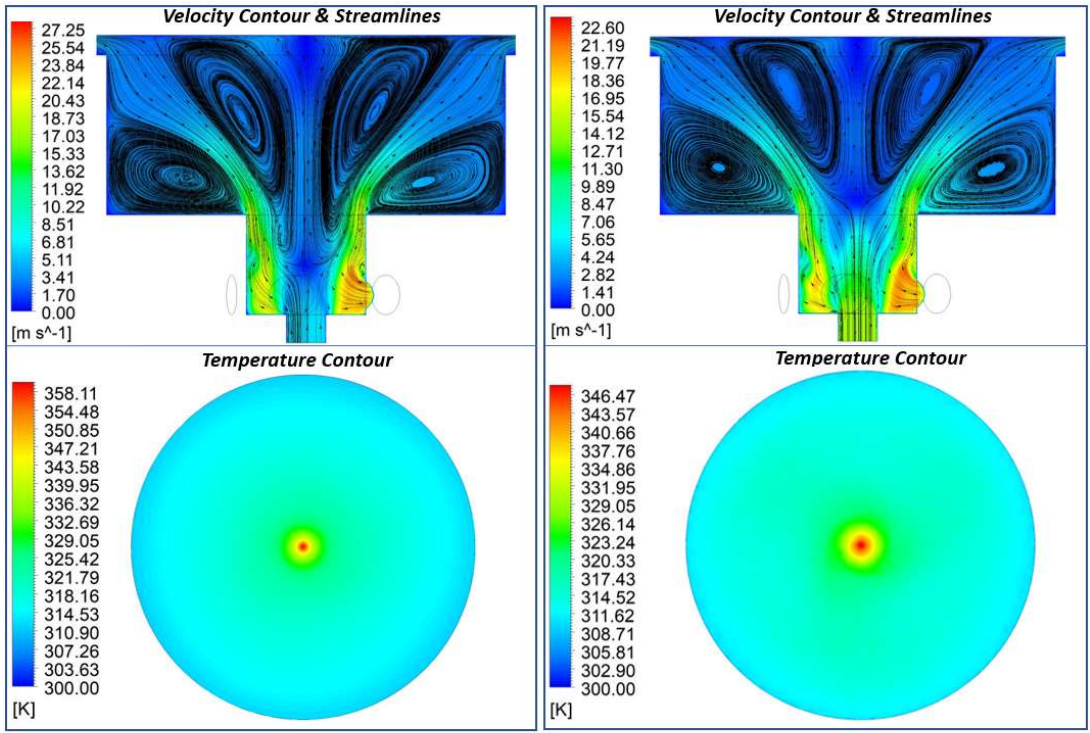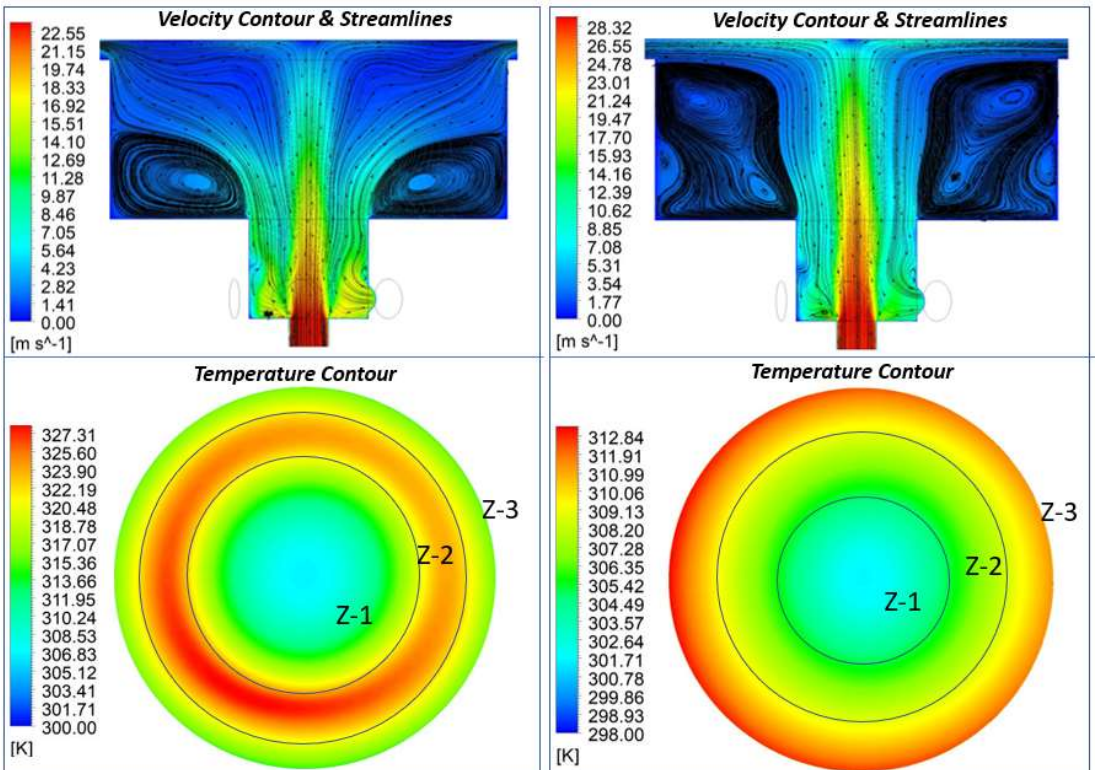

Fig.13. Computed average velocity (with streamlines) and temperature contour plots at all split ratios (a).*SR-1*, (b).*SR-2*, (c).*SR-3*, and (d).*SR-4*.

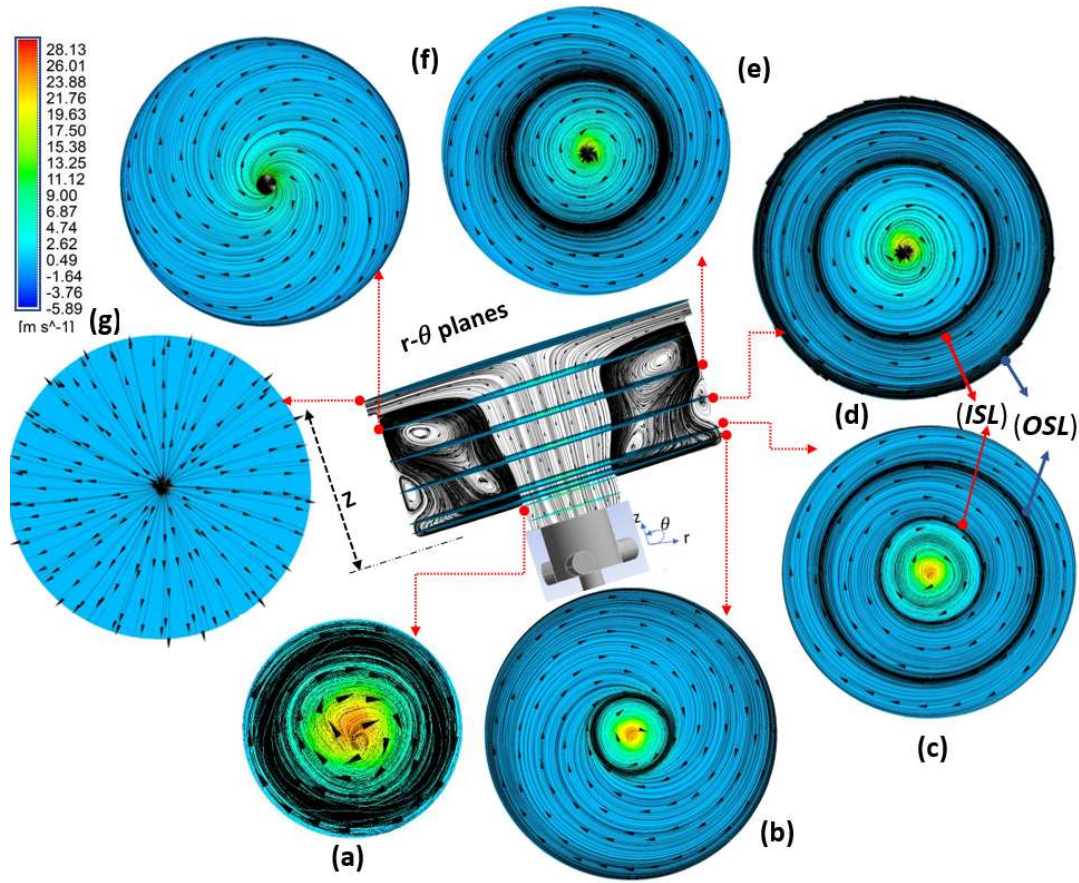

Fig.14. Computed average velocity contours & streamlines at *SR-4* (*Re-12000*) for the proposed augmented jet at various *r-θ* planes *(a) z = 10mm* (from exit plane), *(b) z =15mm, (c) z = 20mm*, *(d) z = 30mm, (e) z = 40mm, (f) z = 50mm* and *(g) z = 60mm* (Impingement plane).

On the other hand, at *SR*-1, which resembles a swirl flow structure (Fig.13. (a)), better heat transfer is at the wall jet regime; more details are very similar to the geometrical swirl jet discussed in section 6.2.

### 6.2 Conventional round jets and swirl jets generated by vane swirlers

Impingement heat transfer simulations for swirling jets generated by geometrical vane swirlers of vane angles *30°, 45°,* and *60°* configurations (refer to section 4), which corresponds to geometrical Swirl number (*S = 1.2, 0.7,* and *0.4*) and conventional round jets both having jet diameter *D = 30mm* are carried out for the Reynolds numbers (*Re - 6500, 9000, 12000,* and *15000*) and at jet-plate impingement distances (*H/D = 1.5, 2, 3,* and *4*) to arrive at optimal conditions for maximum heat transfer rate. Effect of essential parameters such as geometrical swirl no. (*S*) for swirling jet, Reynolds no. (*Re*), and jet-plate distance (*H/D*) are discussed along with flow field (average velocity & streamline) and temperature contour plots.

### 6.2.1. Effect of Swirl no. (S) / vane angle ($\theta$)

The vane or blade angle for a vane-type swirler defines the geometrical swirl number, which quantifies the intensity of the swirl. Fig.15. shows the local Nusselt number distribution for the case $H/D = 2$ for the swirlers at vane angles $60°$, $45°$, and $30°$. Swirler with $60°$ vane angle corresponding to $S = 1.2$ exhibited better heat transfer characteristics with an average Nusselt number ($Nu_{avg} = 48.82$) followed by $30°$ ($S = 0.4$) and $45°$ ($S = 0.7$) with $Nu_{avg} = 48.78$ and $Nu_{avg} = 42.29$ respectively. Unlike round jets, these jets had a minimum heat transfer at the stagnation point ($r/D = 0$), meaning a wide stagnation zone with these swirling jets due to the widening of the jets at the swirler exit [Fig. 18. (b) and Fig. 19. (a)].

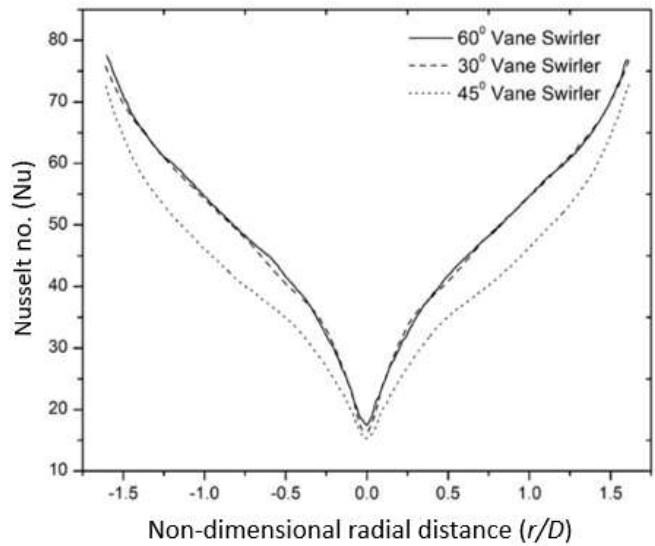

Fig.15. Effect of swirler vane angles ($\theta = 60°$, $30°$, and $45°$) corresponds to geometrical swirl no. ($S = 1.2, 0.4,$ and $0.7$) for the case of $H/D = 2$.

### 6.2.2 Effect of jet-plate distance (H/D)

Fig. 16. shows the effect of jet-plate distance($H/D$) for a conventional round jet and swirling jet generated by a vane swirler, respectively, each of $30mm$ jet diameter. In Fig. 16. (a). for all the cases except at $H/D = 4$, the Nu is minimum at the stagnation point ($r/D = 0$), which indicates that round jets impingement heat transfer is best after a jet-plate distance $H/D = 4$ in par with the literature [2, 3] which says $H/D = 4$-$6$ corresponds to maximum heat transfer as $H/D = 4$ is also the length of potential core for the round jets(ref Section 2). Fig. 17 (a) supports this argument, as we can see a gradual decrement in the width of the stagnation regime with the jet-plate distance ($H/D$).

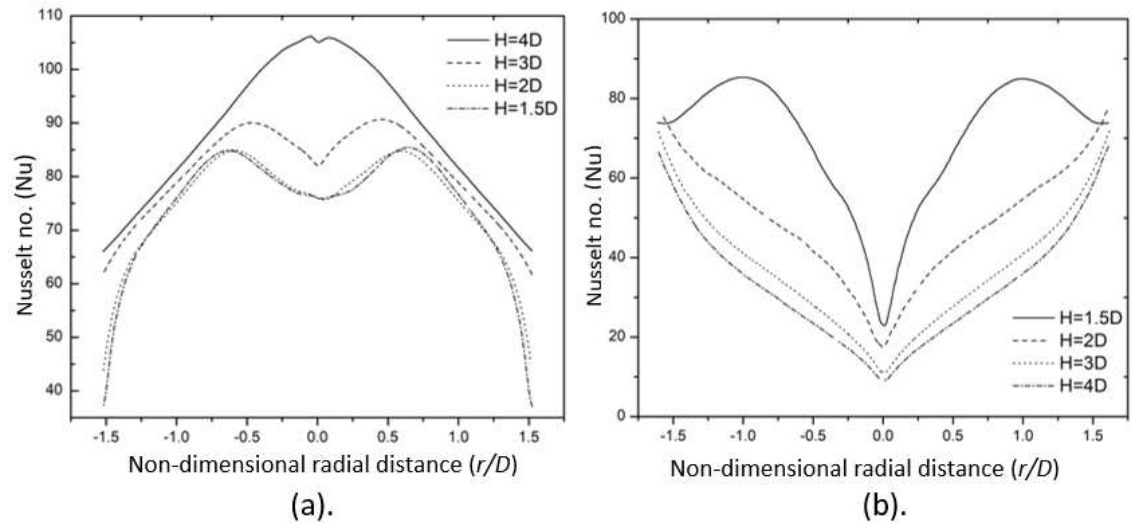

Fig.16. Effect of jet-plate distance on heat transfer at *Re-12000* (a). Round jet (b). *60°* swirler.

Fig.16 (b) shows that for swirling jets generated by the swirler of *60°* vane angle, better heat transfer is at a low jet-plate distance of *H/D = 1.5*, which agrees with swirling jet impingement literature [17,20]. These jets exhibit a good jet spread and heat transfer near the wall jet regimes; however, there is a poor heat transfer characteristics at the stagnation region owing to the nature of the widening of these jets (Fig.17 (b)), which shows an increase in width of stagnation zone with the jet-plate distance(*H/D*). This formed the seed for our idea of augmenting the swirling jets with round jets for impingement heat transfer.

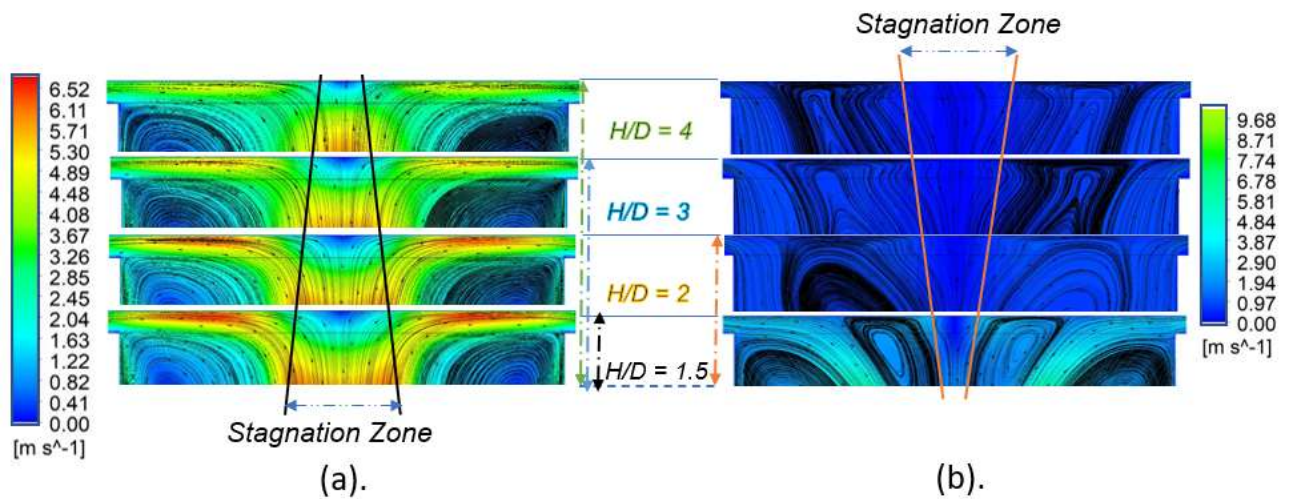

Fig.17. Computed stagnation zones variation with impingement distance (*H/D*) at *Re-12000* (a). Round jet (b). *60°* swirler.

### 6.2.3 Effect of jet Reynolds number (Re)

Fig. 18 (a) shows the effect of inlet jet Reynolds number (*Re*) at *H/D* = 4 (optimized jet-plate distance) for a round jet. It is observed that for all cases except for *Re-12000*, the Nu is maximum at the stagnation point (*r/D* = 0), and it decreases along the impingement regime and wall jet regimes (refer to Fig 19. (b)). For the *Re-12000* case, the Nu is minimum at the stagnation point(*r/D=0*) and increases up to the impingement regime ($0 \geq r/D \geq 0.5$) and again decreases along the wall jet regime ($0.5 \geq r/D \geq 1.6$). Fig. 18 (b) shows the effect of inlet jet Reynolds number (*Re*) at *H/D* = 1.5(optimized jet-plate distance *H/D*) for a swirling jet. For all the cases, the *Nu* is minimum at the stagnation point (*r/D* = 0) and increases steadily along the impingement regime ($0 \geq r/D \geq 1.0$) and tries to stabilize near the wall jet regime ($1.0 \geq r/D \geq 1.6$).

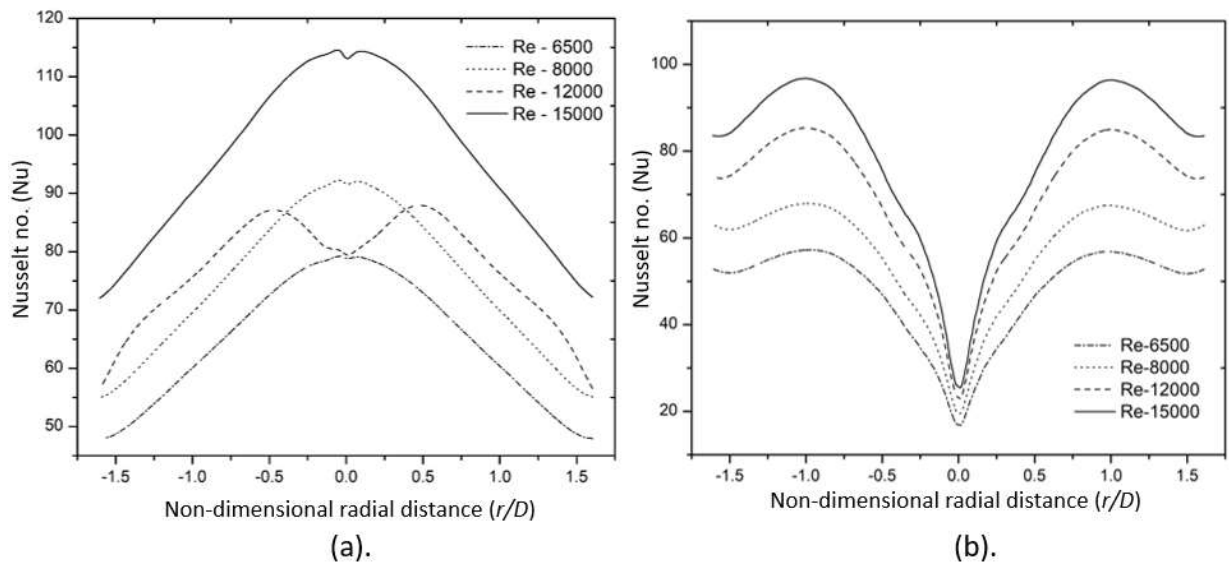

Fig.18. Effect of jet Reynold number for the case of (a). Round jet at *H/D* = 4. (b). *60°* vane swirler jet at *H/D* = 1.5

### 6.2.4 Flow field and temperature contours

Fig. 19. (a) shows the computed average velocity contour and streamlines plot for a round and *60°* vane swirling jet impingement. In compliance with Fig.1, 2 (a) & (b). the potential core, shear layers, and different regimes of impinging jets viz stagnation, impingement, and wall jet regimes are indicated (refer Fig. 1, 2). Fig. 19 (b) shows the computed temperature contours for the jets at *Re-12000*. In the figure, *Z-1, Z-2, and Z-3* correspond to stagnation, impingement, and wall jet regimes. Fig.20 shows the computed averaged velocity contours and streamlines for the direct jet at various *r-θ* planes.

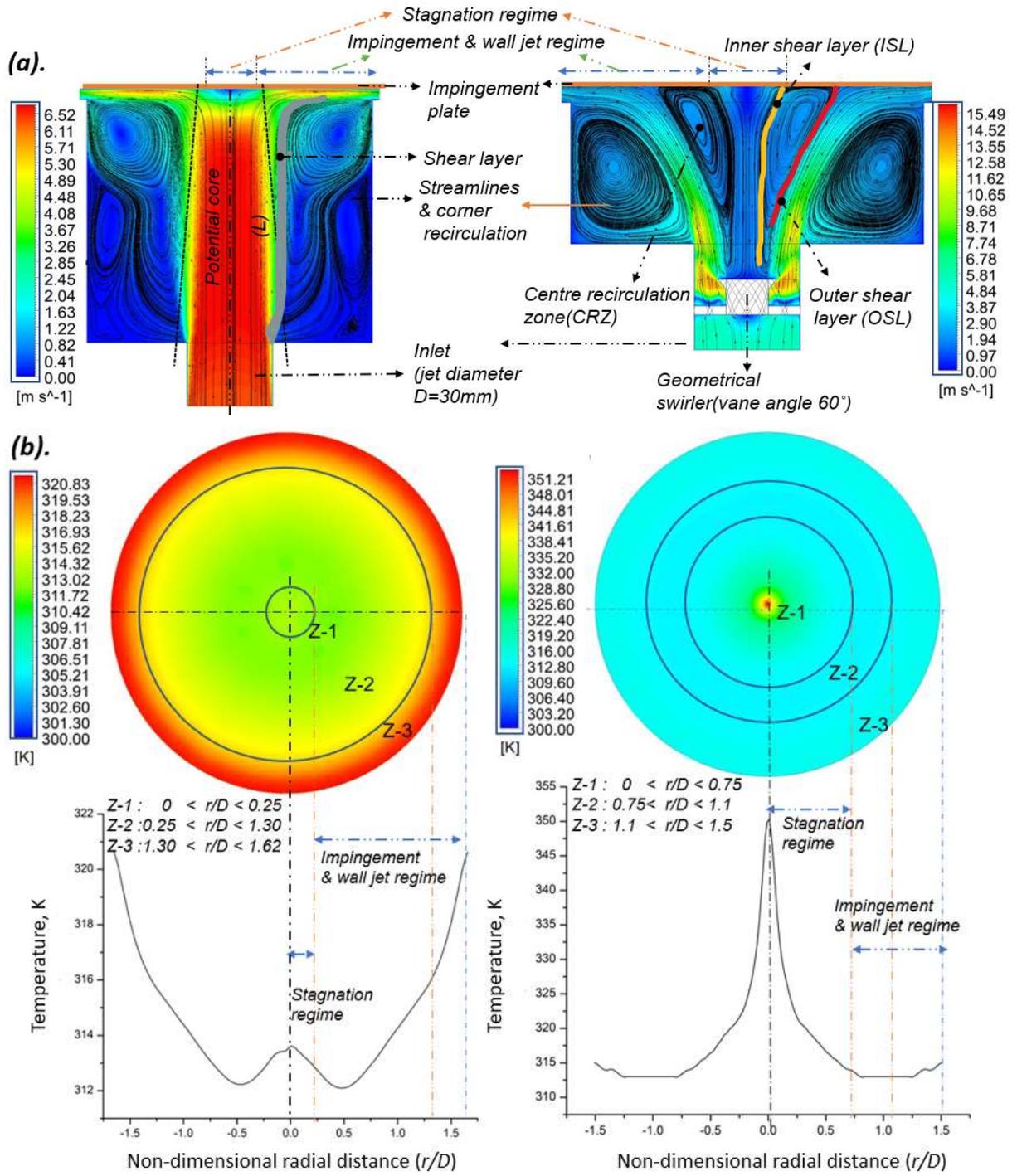

Fig.19. Computed contour plots for round jet at *H/D = 4* and *60˚* vane swirler jet at H/D = *1.5* (a). Average velocity and streamline at an orthogonal front (*r-z*) plane (b). Temperature contour on the impingement (*r- θ*) plane at *H/D = 4* for round jet and *H/D = 1.5* for *60˚* vane swirler jet.

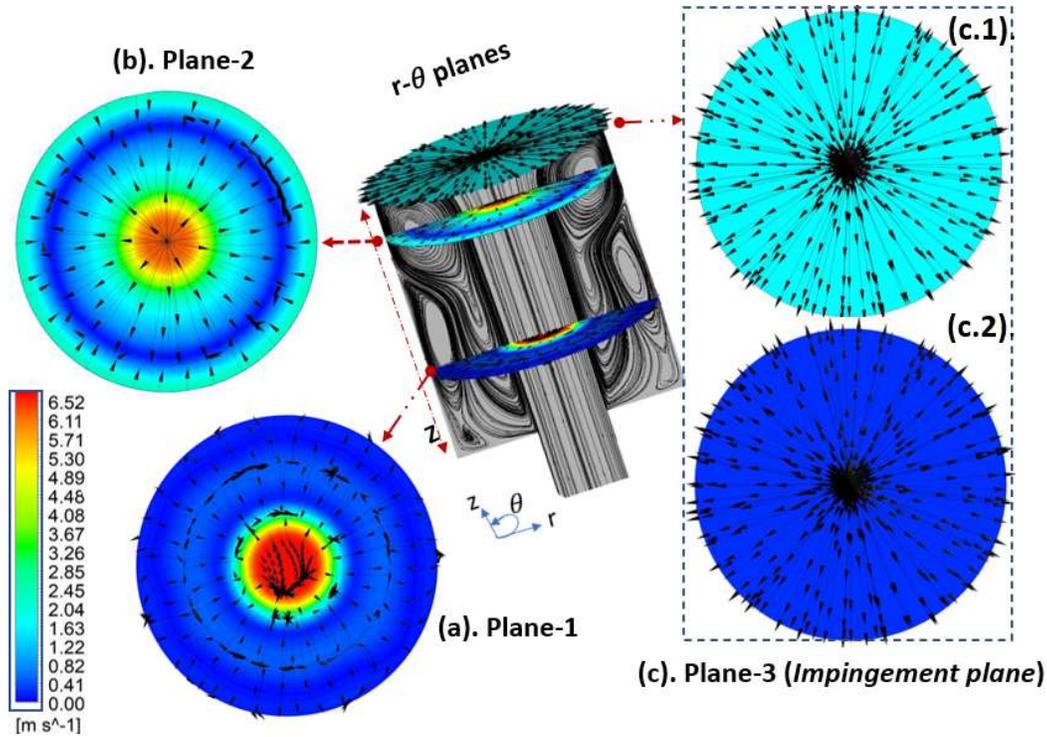

Fig.20. Computed average velocity contours & streamlines for a round jet at r-θ planes *(a) z = 50mm* (from the exit), *(b) z = 110mm*, and *(c) z =150mm* (Impingement plane).

At *z =50 mm*, there is a dominance of axial velocity centered near the jet core. At *z =110 m*m, there should be entrainment of surrounding air to the jet, which makes the streamlines look like radial outward lines from the center and radial inward lines at the periphery. At the impingement plane *z =150mm*, the flow is strictly directed radially outwards with a stagnation core at the center. In Fig.20, (c.1) indicates the dominance of the radial velocity component, and (c.2) shows a zero contour of axial velocity component at the impingement plane as there is a change in direction vector from axial to the radial due to the presence of impingement plate. Similarly, Fig.21 shows the computed averaged velocity contours and streamlines for a vane swirler of *60°* at various r-θ planes. There is a strong swirl from *z = 10mm* to *Z = 55mm*. At the impingement plane *z = 55mm*, the streamlines are swirl-radially outward beyond the vortex region and swirl-radially inward in the vortex region demarcated by the inner shear layer (ISL) (refer Fig 21. (f) *z =55mm*), which separates the vortex flow (vortex-induced re-circulation zones) from the mean flow field. At *z = 70mm*, there is no swirl component, as the streamlines indicate circular lines meaning a non-swirl domain, which supports the argument of the dominance of swirl flow [17,20,24] at a low jet-plate distance (*H/D = 1.5*).

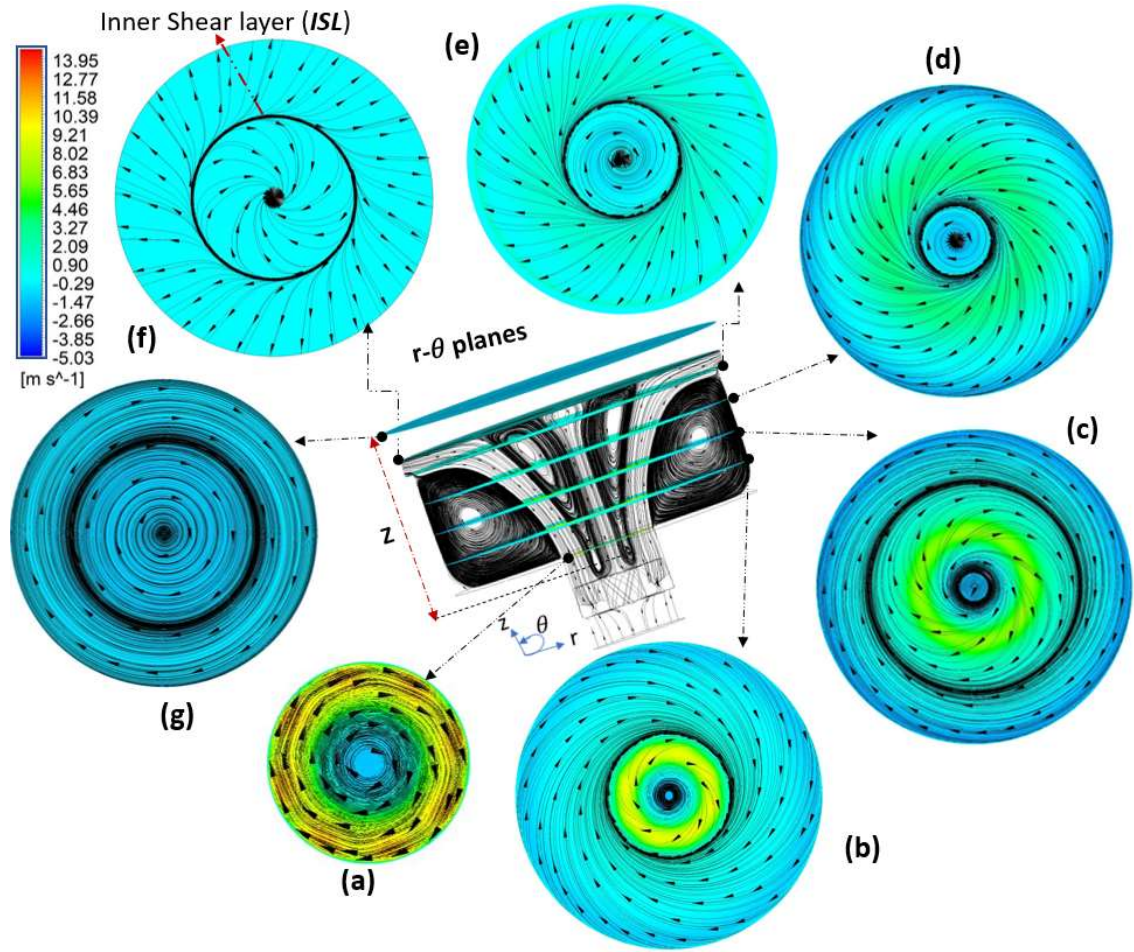

Fig.21. Computed average velocity contours & streamlines for a 60° vane swirler jet at various *r- θ* planes *(a) z = 10mm* (swirler dump plane), *(b) z = 20mm*, *(c) z =30mm*, *(d) z = 40mm*, *(e) z = 50mm, (f) z = 55mm* (Impingement plane), and *(g) z = 70mm* (non-swirl domain).

### 6.3 Comparative analysis

Sections 6.1 and 6.2 dealt with the parametric study with crucial parameters that dictates the impingement heat transfer besides discussing the flow structures for all the jet types. It is inferred that optimized *H/D = 4* for round jets and *H/D = 1.5* for both the proposed augmented and vane swirler jets. Also, a Swirl no. (*S = 1.2*) for a 60°vane swirler jet and a Split ratio of *SR-4* split ratio for augmented jet are found to be other optimum conditions. This section presents a comparative analysis of heat transfer quantified in terms of local Nusselt no. (*Nu*) distribution, average Nusselt no. (*Nu avg*) and stagnation Nusselt no. (*Nu stg*) for all the jets at their optimum parameters. Also, the average turbulence kinetic energy (*TKE*) at a minimum and maximum *H/D* for all the jets are compared.

### 6.3.1 Comparative analysis of heat transfer at optimal conditions for all jets

The local Nusselt no. (*Nu*) distribution at optimum conditions for maximum heat transfer for the proposed augmented jet (at *SR-4* and *H/D = 1.5*), Round jet (at *H/D = 4*), and vane swirler jet (at *S = 1.2* (or) $\theta = 60^0$ and *H/D = 1.5*) at all the Reynolds no. (*Re - 6500, 8000, 12000,* and *15000*) are presented in Fig.22.

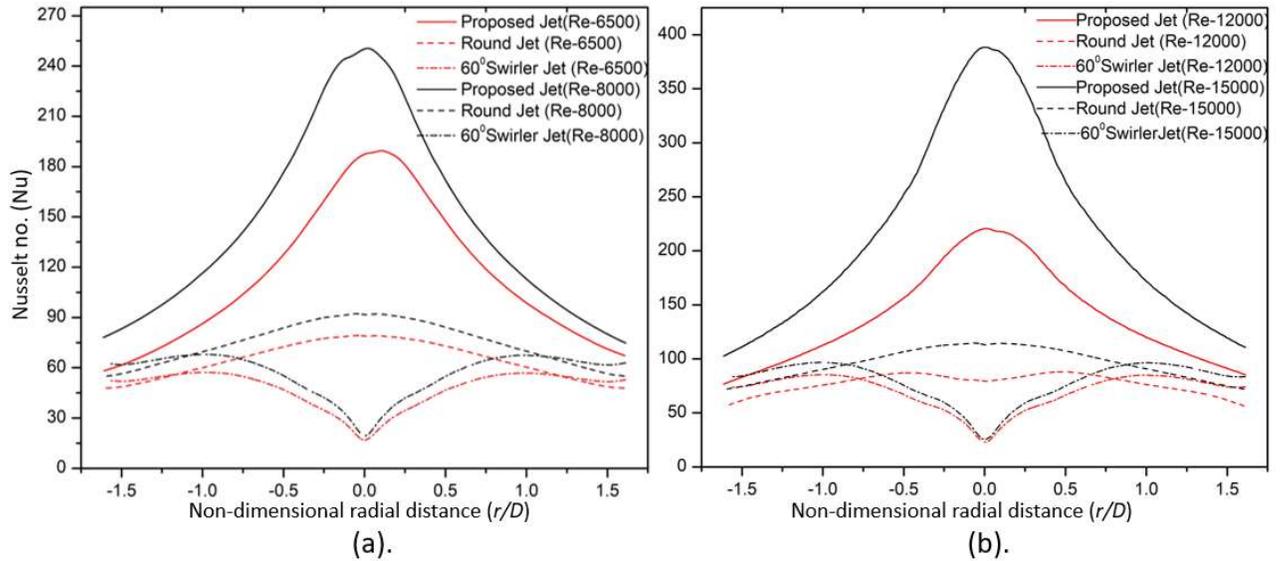

Fig.22. Nusselt number distribution for all the jet types at optimum conditions of jet-plate distance (*H/D*) and Split ratio (*SR*)/ Swirl no. (*S*). (a) *Re-6500, 8000* (b). *Re- 12000, 15000*.

The average Nusselt no. (*Nu $_{avg}$*) and stagnation Nusselt no. (*Nu $_{stg}$*) at different jet-plate distances (*H/D*) for all the jets are shown in Fig.24. The effect of Reynolds no. (*Re*) and jet-plate distance (*H/D*) on the average and stagnation Nusselt no. is shown in Fig.25. and Fig.26. respectively. For the proposed augmented jet, the *Nu $_{avg}$* and *Nu $_{stg}$* are maximum at a low jet-plate distance (*H/D =1.5*), gradually decreases up to *H/D =3* and again increases up to *H/D =4*. A similar trend happens with Reynolds no (*Re*) for the proposed jet. For a round jet, *Nu $_{avg}$* and *Nu $_{stg}$* increase with jet-plate distance (*H/D*) and Reynolds no. (*Re*) on par with the literature [3]. For the vane swirler jet, *Nu $_{avg}$* increases linearly with Reynolds no. (*Re*) and decreases with *H/D*, whereas *Nu $_{stg}$* increases between *Re = 6500 - 8000*, decreases between *Re = 8000 - 12000*, and again increases up to *Re = 15000*. A similar trend in *Nu $_{stg}$* happens with *H/D* for these jets.

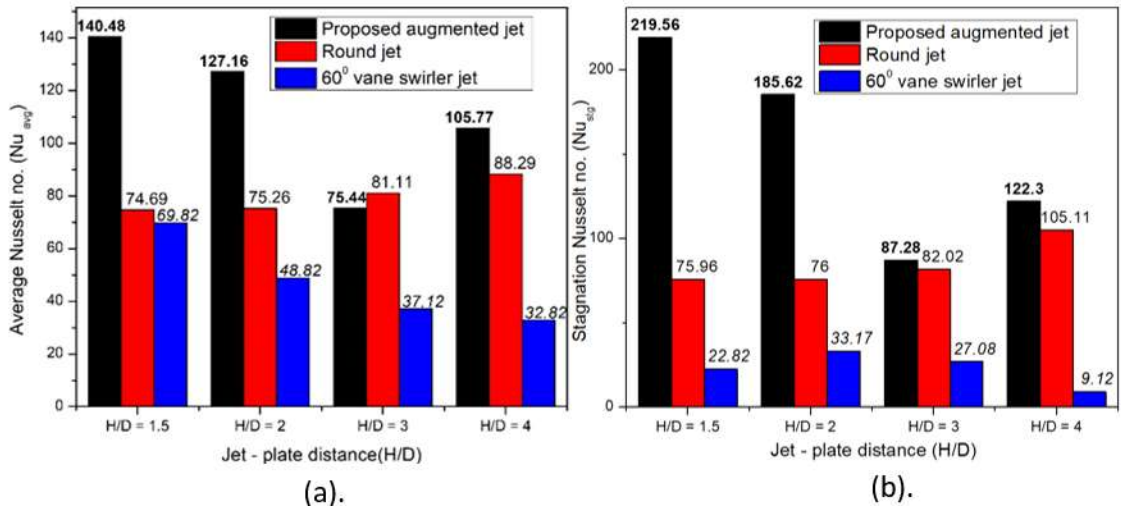

Fig.24. Comparison of Nusselt no. for all jets at different *H/D* (a). Average Nusselt no. (*Nu avg*) (b). Stagnation Nusselt no. (*Nu stg*).

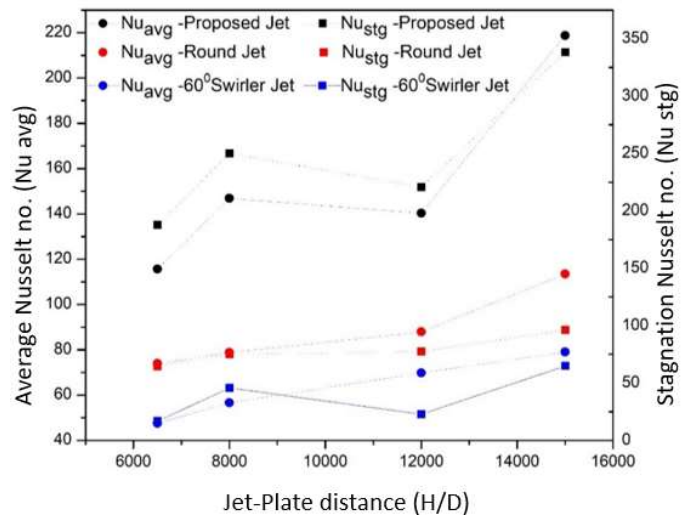

Fig.25. Effect of Reynolds number on heat transfer for all jets quantified in terms of average and stagnation Nusselt number at optimal conditions of jet-plate distance (*H/D*) and Split ratio (*SR*)/ Swirl no. (*S*).

### 6.3.2 Comparison of turbulence kinetic energy (TKE) for all jets

Fig. 25. shows the contours of turbulence kinetic energy at the minimum and maximum jet-plate impingement distances *H/D =1.5* and *H/D =4* at the optimal conditions of Split ratio *SR-4* for the proposed augmented jet and *S=1.2* for the swirl jet for Reynolds number (*Re- 12000*). It is evident from the plot that turbulence kinetic energy (*TKE*) is maximum at a low jet-plate distance *H/D = 1.5* for the proposed augmented jet and swirl jet, on the other hand,

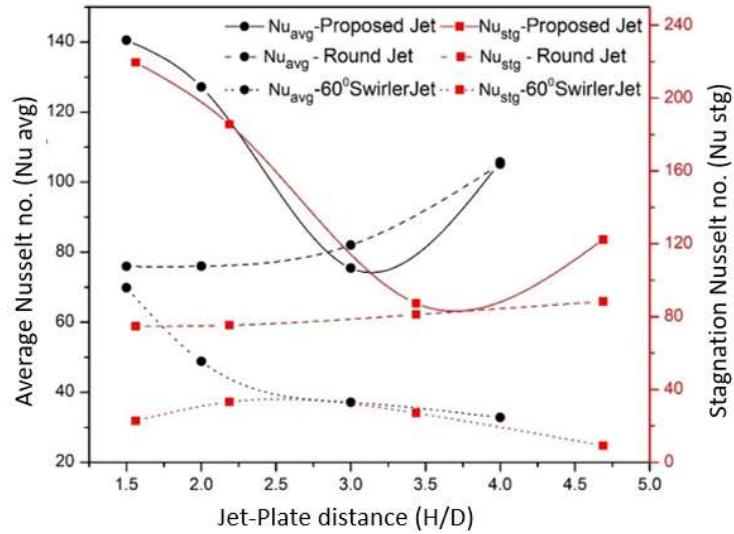

Fig.26. Effect of jet-plate distance (*H/D*) on heat transfer for all jets quantified in terms of average and stagnation Nusselt number at Reynolds no. (*Re-12000*).

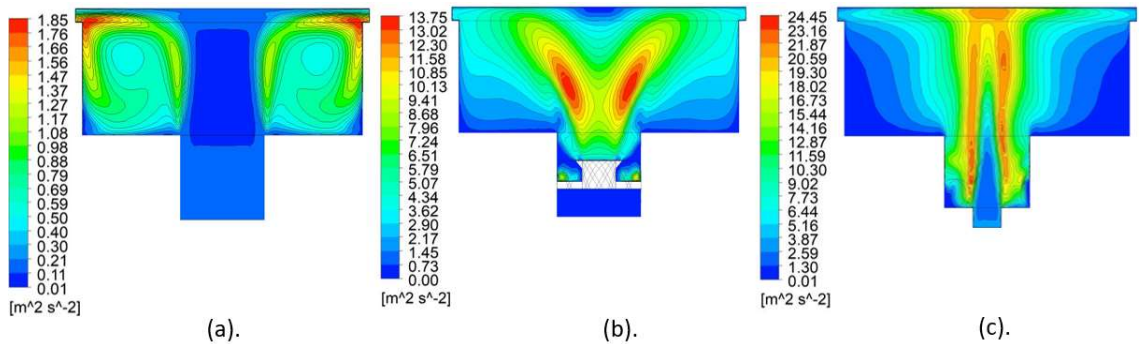

Fig. 27. Comparison of turbulence kinetic energy contour at jet-plate distances *H/D = 1.5* at *Re-12000* for (a). Round jet and (b). Swirling jet by vane swirler (c). Proposed augmented jet.

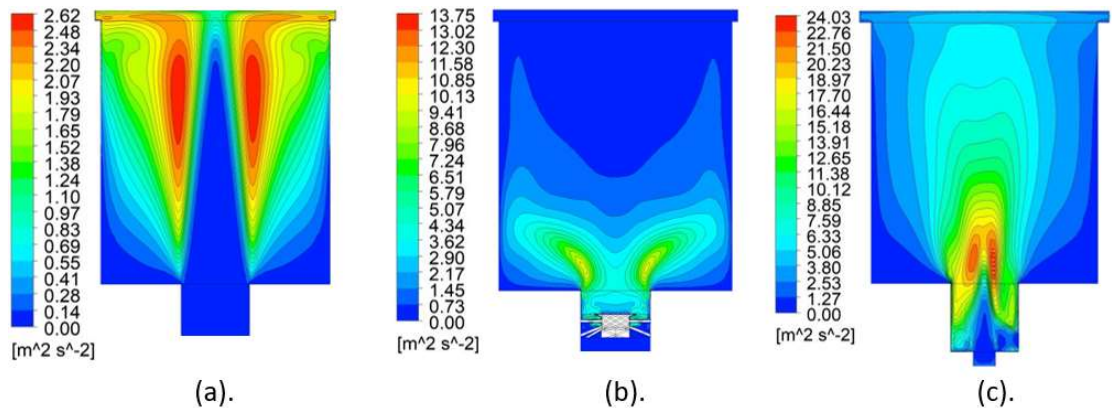

Fig. 28. Comparison of turbulence kinetic energy contour at jet-plate distances *H/D = 4* at *Re-12000* for (a). Round jet and (b). Swirling jet by vane swirler (c). Proposed augmented jet.

at $H/D = 4$ for the conventional round jet, which eventually enhances heat transfer at these impingement distances as *TKE* quantifies the intensity of turbulence. Also, the magnitude of this quantity, i.e., TKE for the proposed jet, is about an order higher than the conventional round jet at a low jet-plate distance ($H/D = 1.5$) which is the reason for better heat transfer.

## 7. Conclusion

The RANS turbulence models are used for solving jet impingement flow and heat transfer problems for the proposed augmented (combined swirling and round) jet by systematically validating the experimental swirl (velocity components) with Gopakumar et al. [4] and impingement heat transfer with Baughn et al. [2]. Also, the numerical simulations for round jets and swirling jets generated by the geometrical vane swirler are performed. The effect of Reynolds number (*Re*), jet–plate distance (*H/D*), and geometrical Swirl number (*S*) for the conventional round and swirling jets are studied. A parameter called split ratio is introduced for the proposed augmented jet, which defines the percentage of flow split through the *3* tangential inlets and an axial inlet port. Thus, the proposed augmented jet generation mechanism can also be operated in hybrid modes to generate a purely swirl jet using the *3* tangential inlets or generate a purely round jet using an axial inlet independently.

From the parametric study, the heat transfer enhancement was found to be a direct function of inlet Reynolds number (*Re*), optimized jet–plate distance *H/D =4* (round jets) and *H/D = 1.5* for the swirl and augmented jets, geometrical Swirl number ($S = 1.2$ or $\theta = 60^0$), and split ratio (*SR-4* for the augmented jet). A comparative study of the flow field using velocity contour and streamline plots are presented for all the jets, along with the results from a limited PIV experimental flow visualization study.

At split ratios *SR-1* and *SR-2*, the flow resembles typical swirling structures for the proposed jet. At higher split ratios *SR-3* and *SR-4*, the percentage of flow through central jet counterparts increases, thus resulting in distortion of re-circulation zones, and the resulting flow feature is an augmented swirl and round jet. At lower jet-plate impingement distance $H/D = 1.5$, the proposed augmented jet and vane swirler jets exhibit better heat transfer characteristics. Increasing the impingement distance has a negative effect on heat transfer for these jets on par with the literature [17,20]. Instead, the round jets exhibit an enhanced heat transfer [3] at a higher jet-plate impingement distance $H/D = 4$. Also, the intensity of turbulence, quantified in terms of turbulence kinetic energy (TKE), is higher at the low jet-plate distance ($H/D = 1.5$) for the proposed augmented jet and swirling jets.

From the comparative heat transfer analysis for all the jets (refer Fig.24 (a)), the stagnation Nusselt number ($Nu_{stg}$) at $r/D = 0$ for the proposed augmented jet at an optimized $H/D = 1.5$ and at split ratio *SR-4*, which has been predicted to be $Nu_{stg} = 220$ which is *189%* higher than the round jet. Similarly, the average Nusselt number ($Nu_{avg}$), which accounts for an averaged heat transfer distribution on the impingement plate, is *140.48* for the proposed jet, which is *88%* higher than the round jet and *101%* higher than the 60° vane swirler jet. The stagnation Nusselt no. ($Nu_{stg}$) and average Nusselt no. ($Nu_{avg}$) for round jets are maximum at $H/D = 4$. The swirling jets by the vane swirler exhibits a poor average and stagnation Nusselt no., which is attributed due to the widening of jets resulting in a flow deficit region near the stagnation region (refer Fig. 24, Fig. 17 (b)).

The maximum heat transfer characteristics are near the stagnation and impingement regimes ($0 \geq r/D \geq 1.30$) at split ratios SR-4, like a round jet profile for heat transfer. Whereas, at lower split ratios SR-1 and SR-2, an enhanced heat transfer at the wall jet regime ($1.30 \geq r/D \geq 1.60$) is like a swirling jet. Hence, operating the proposed augmented jets at intermittent modes of split ratio is recommended to ensure a uniform heat transfer for practical heat transfer/cooling applications. Hence, the proposed augmented jet, which enhances heat transfer from the present predictive numerical analysis, can be a suitable candidate for any jet impingement heat transfer applications; however, an experiment using temperature measurements is required for verifying the numerical results reserved for future work.

**Acknowledgement:**

The first author wants to acknowledge Dr. Gopakumar (Sandia National Laboratories, Livermore, California) for sharing the vane swirler geometry details used for his combustor flame stabilization studies which is also used in the present study.

**Appendix A. Turbulence modelling using RANS**

In the present work, numerical simulations are carried out using Transition $k$ - $k_l$ – $\omega$ and *SST* $k - \omega$ *RANS* turbulence models. The former showed a good predictive capability for both swirling and non-swirling jet impingement heat transfer at low jet–plate distance $H \leq 2D$, whereas the latter was compatible with predicting jet impingement at a low jet-plate distance i.e., $H > 2D$.

**Appendix A.1. Transition $k$ - $k_l$ - $\omega$ model**

The transport equations for turbulent kinetic energy ($k_T$), laminar kinetic energy ($k_L$) and specific dissipation rate ($\omega$) are

$$\frac{Dk_T}{Dt} = P_{k_T} + R + R_{NAT} - \omega\, k_T - D_T + \frac{\partial}{\partial x_j}\left[\left(v + \frac{\alpha_T}{\alpha_k}\right)\frac{\partial k_T}{\partial x_j}\right]$$

(A.1)

$$\frac{Dk_L}{Dt} = P_{k_L} - R - R_{NAT} - D_L + \frac{\partial}{\partial x_j}\left[v\frac{\partial k_L}{\partial x_j}\right]$$

(A.2)

$$\frac{D\omega}{Dt} = C_{\omega 1}\frac{\omega}{k_T}P_{k_T} + \left(\frac{C_{\omega R}}{f_W} - 1\right)\frac{\omega}{k_T}(R + R_{NAT}) - C_{\omega 2}\omega^2 + C_{\omega 3}f_\omega \alpha_T f_W^2 \frac{\sqrt{K_T}}{d^3}$$

$$+ \frac{\partial}{\partial x_j}\left[\left(v + \frac{\alpha_T}{\alpha_k}\right)\frac{\partial \omega}{\partial x_j}\right]$$

(A.3)

Where, $P_{k_T}$ and $P_{k_L}$ in the equations (A.1), (A.2), and (A.3) are the terms for production of turbulent and laminar kinetic energy respectively generated by their corresponding small-scale turbulent viscosity ($v_{T,L}$) and large-scale turbulent viscosity ($v_{T,S}$). $R$ represents the averaged effect of streamwise fluctuations on turbulence during bypass transition. $R_{NAT}$ is the natural transition production term arising due to instabilities breaking down laminar to a turbulent flow [27, 17]. The ratio of effective length scale to the turbulent length scale is given by the $f_W$.

The eddy viscosity and thermal viscosity are modelled as given below,

$$\overline{-\acute{u}_i\acute{u}_j} = v_{TOT}\left(\frac{\partial u_i}{\partial x_j} + \frac{\partial u_j}{\partial x_i}\right) - \frac{2}{3}k_{TOT}\delta_{ij} \qquad (A.4)$$

$$\overline{-\acute{u}_iT'} = \alpha_{\theta,TOT}\frac{\partial \bar{T}}{\partial x_i} \qquad (A.5)$$

Where, the total eddy viscosity ($v_{TOT}$) is the sum of large – scale turbulent viscosity ($v_{T,l}$) and small – scale turbulent viscosity ($v_{T,S}$) by,

$$v_{TOT} = v_{T,S} + v_{T,l}$$

(A.6)

The total eddy diffusivity ($\alpha_{\theta,TOT}$) is given by,

$$\alpha_{\theta,TOT} = f_W \left(\frac{k_T}{k_{TOT}}\right) \frac{v_{T,s}}{Pr_\theta} + (1 - f_W)C_{\alpha,\theta} \sqrt{k_T} \lambda_{eff} \quad (A.7)$$

The turbulent scalar diffusivity ($\alpha_T$) and total kinetic energy ($k_{TOT}$) in the equations (A.3) and (A.4) are given by,

$$\alpha_T = f_V \left(\frac{k_T}{k_{TOT}}\right) C_{\mu,std} \sqrt{k_{T,s}} \lambda_{eff} \quad (A.8)$$

$$k_{TOT} = k_T + k_L \quad (A.9)$$

The other details of Transition $k$ - $k_l$ - $\omega$ model are in Ansys fluent theory guide [23]. Following are the model constants used in the numerical simulation:

$C_\mu = 0.09, C_\lambda = 2.495, C_R = 0.12, A_{NAT} = 200, A_{TS} = 200, C_{NAT,crit} = 1250, CR_{NAT} = 0.02, A_v = 2.495, C_{INT} = 0.75, C_{\omega 1} = 0.44, C_{\omega 3} = 0.3, C_{\alpha\theta} = 0.035, C_{\tau l} = 4360, Pr_{TKE} = 1, Pr_{SDR} = 1.17, Pr_{energy} = 0.85, Pr_{wall} = 0.85$.

**Appendix A.2. $k - \omega$ (SST) model**

The model is a blend of standard $k - \omega$ model in the vicinity of wall to a modified $k$ -$\varepsilon$ model away from the wall [11,23]. The transport equations for $k$ and $\omega$ are,

$$\frac{\partial}{\partial t}(\rho k) + \frac{\partial}{\partial x_i}(\rho k u_i) = \frac{\partial}{\partial x_j}\left[\left(\mu + \frac{\mu_t}{\sigma_k}\right)\frac{\partial k}{\partial x_j}\right] + min(P_k, 10\rho\beta^* k\omega) - \rho\beta^* k\omega$$

(A.10)

$$\frac{\partial}{\partial t}(\rho\omega) + \frac{\partial}{\partial x_i}(\rho\omega u_i) = \frac{\partial}{\partial x_j}\left[\left(\mu + \frac{\mu_t}{\sigma_k}\right)\frac{\partial \omega}{\partial x_j}\right] + \alpha \frac{\omega}{k} P_k - \rho\beta\omega^2 + 2(1 - F_1)\frac{\rho\sigma_{\omega 2}}{\omega}\frac{\partial k}{\partial x_j} \cdot \frac{\partial \omega}{\partial x_j}$$

(A.11)

The last term in the right side of the eq. (A.11) is the cross-diffusion term. The coefficients are,

$$\sigma_k = \frac{1}{\frac{F_1}{\sigma_{k,1}} + \frac{(1 - F_1)}{\sigma_{k,2}}}$$

(A.12)

$$\sigma_k = \frac{1}{\frac{F_1}{\sigma_{\omega,1}} + \frac{(1-F_1)}{\sigma_{\omega,2}}}$$

(A.13)

$$\alpha_\infty = F_1 \alpha_{\infty,1} + (1-F_1)\alpha_{\infty,2} \qquad (A.14)$$

where, $F_1$ and $F_2$ are the blending functions. The turbulent viscosity is modelled using,

$$\mu_t = \rho_{C_\mu}(\bar{v})^2 \frac{k}{\varepsilon} \qquad (A.15)$$

The other details of *SST k - ω* model are in Ansys fluent theory guide [23]. Following are the model constants used in the numerical simulation: $\alpha_\infty^* = 1$, $\alpha_\infty = 0.52$, $\beta_\infty^* = 0.09$, $a_1 = 0.31$, $Pr_{TKE} = 1$, $\beta_{i,1} = 0.075$, $\beta_{i,2} = 0.0828$, $\sigma_{k,1} = 1.176$, $\sigma_{k,2} = 1$, $\sigma_{\omega,1} = 2$, $\sigma_{\omega,2} = 1.168$, $Pr_{energ} = 0.85$, $Pr_{wall} = 0.85$.